\documentclass[epj]{svjour}
\usepackage{graphics}
\begin{document}
%
\title 
{
On the Sensitivity of L/E Analysis of Super-Kamiokande Atmospheric Neutrino Data to Neutrino Oscillation  Part 1
}
\subtitle
{--- The Effect of Quasi-Elastic Scattering over the Direction of the Emitted Lepton in  the Neutrino Events inside the Super-Kamiokande Detector ---}

\author{E. Konishi\inst{1}, Y. Minorikawa\inst{2}, V.I. Galkin\inst{3},
M. Ishiwata\inst{4}, I. Nakamura\inst{4}, N. Takahashi\inst{1}, M. Kato\inst{5}  \and A. Misaki\inst{6}\inst{7}    
}
\institute{
Graduate School of Science and Technology, Hirosaki University, Hirosaki, 036-8561, Japan    
\and Department of Science, School of Science and Engineering, Kinki University, Higashi-Osaka, 577-8502, Japan
\and Department of Physics, Moscow State University, Moscow, 119992, Russia
\and Department of Physics, Saitama University, Saitama, 338-8570, Japan
\and Kyowa Interface Science Co.,Ltd., Saitama, 351-0033, Japan
\and Inovative Research Organization, Saitama University, Saitama,
 338-8570, Japan
\and Research Institute for Science and Engineering, Waseda University, Tokyo, 169-0092, Japan 
\\\email{konish@si.hirosaki-u.ac.jp}
}

\abstract {
It is said that the finding of the maximum oscillation in neutrino 
oscillation by Super-Kamiokande is one of the major achievements of the 
SK.
 In present paper, we examine the assumption made by Super-Kamiokande
 Collaboration that the direction of the incident neutrino is
 approximately the same as that of the produced lepton,
which is the cornerstone in their $L/E$ analysis
 and we find this approximation does not hold even approximately.
In the Part 2 of the subsequent paper, we apply the results from 
Figures 12, 13 and 14 to $L/E$ analysis and conclude that one cannot
 obtain the maximum oscillation in $L/E$ analysis which shows strongly
 the oscillation pattern from the neutrino oscillation.
}

\authorrunning{E.Konishi et. al.,}
\titlerunning{On the Sensitivity of L/E Analysis of SK Neutrino Oscillation}
\maketitle
%

\section{Introduction}
According to the results obtained from the Super-Kamio-\
kande Experiments on atmospheric neutrinos, 
it is said that oscillation phenomena have been found between 
two neutrinos, $\nu_{\mu}$and $\nu_{\tau}$.
 Published reports on the confirmation to the oscillation between the 
neutrinos, $\nu_{\mu}$and $\nu_{\tau}$, and the history foregoing 
these experiments will be 
critically reviewed and details are in the following:  

\begin{itemize}
\item[(1)]During 1980's Kamiokande and IMB observed smaller atmospheric 
neutrino flux ratio $\nu_{\mu}/\nu_e$ than the predicted value
\cite{Hirata}.
\item[(2)]Kamiokande found anomaly in the zenith angle distribution \cite{Hatakeyama}.
\item[(3)]Super-Kamiokande found $\nu_{\mu}$-$\nu_{\tau}$ oscillation 
\cite{Kajita2} and \\
Soudan2 and MACRO confirmed the Super-Kamiokande result
\cite{Mann}.
\item[(4)]K2K, the first accelerator-based long baseline experiment, 
confirmed atmospheric neutrino oscillation\cite{K2K}.
\item[(5)]MINOS's precision measurement gives the consistent results
with Super-Kamiokande ones\cite{MINOS}.
\end{itemize}

  It is well known 
that Super-Kamiokande Collaboration
examined all possible types of the
neutrino events, such as, say, Sub-GeV e-like, Multi-GeV e-like,
Sub-GeV $\mu$-like, Multi-GeV $\mu$-like, Multi-ring Sub-GeV $\mu$-like,
 Multi-ring 
Multi-GeV $\mu$-like, PC, {\it Upward Stopping Muon Events} and 
{\it Upward Through Going Muon Events}.
In other words, all possible interactions by neutrinos, such as, 
elastic and quasielastic scattering, single-meson production and
deep scattering are considered in their analyses.
 Furthermore, all topologically 
different types of neutrino events lead to the unified numerical 
oscillation parameters, say,  
$\Delta m^2 = 2.4\times 10^{-3}\rm{ eV^2}$ and $sin^2 2\theta=1.0$
\cite{Ashie2}.\
 
However, these parameters are obtained from the analysis of
the zenith angle distribution 
of the incident neutrinos which are based on the survival probability 
of a given flavor( see Eq.(7)).
The most important result among the achievements on neutrino oscillation
 made by Super-\\
Kamiokande Collaboration is the finding of the maximum oscillation 
in neutrino oscillation, because it is the ultimate result
in the sense that they observe the oscillation pattern itself
directly in neutrino oscillation.

  Taking account of all factors mentioned above, it is natural
that the majority believes the finding of the $\mu-\tau$ 
neutrino oscillation by Super-Kamiokande Collaboration.
 
  However, it should be emphasized strongly that 
Super-Kamiokande Collaboration put the fundamental assumption in 
all possible analyses of the atmospheric neutrino
oscillation which is never self-evident and should be carefully examined.
 This assumption is that the directions of the incident 
neutrinos are approximately the same as those of emitted leptons.

In order to avoid any misunderstanding toward 
the SK assumption on the direction, we reproduce this assumption 
from the original SK papers and their related papers in italic.

[A] Kajita and Totsuka \cite{Kajita1} state
\footnote{see page 101 of the paper concerned.}:
\begin{quote}
"{\it However, the direction of the neutrino must be estimated from the
reconstructed direction of the products of the neutrino interaction.
 In water Cheren-kov detectors, the direction of an observed lepton is
assumed to be the direction of the neutrino. Fig.11 and 
Fig.12 show the estimated correlation angle between 
neutrinos and leptons as a function of lepton momentum.
 At energies below 400~MeV/c, the lepton direction has little 
correlation with the neutrino direction. The correlation angle 
becomes smaller with increasing lepton momentum. Therefore, 
the zenith angle dependence of the flux as a consequence of 
neutrino oscillation is largely washed out below~400 MeV/c lepton
momentum. With increasing momentum, the effect can be seen more 
clearly.}" 
\end{quote}

[B] Ishitsuka \cite{Ishitsuka} states\footnote
{see page 138 of the paper concerned.}:
\begin{quote}
" {\it 8.4  Reconstruction of $L_\nu$
\vskip 2mm

Flight length of neutrino is determined from the neutrino incident
zenith angle, although the energy and the flavor are also involved.
 First, the direction of neutrino is estimated for each sample by 
a different way. Then, the neutrino flight lenght is 
calclulated from the zenith angle of the reconstructed direction.
\\
\\
 8.4.1 Reconstruction of Neutrino Direction

\vspace{-2mm}

{\flushleft{\underline 
{FC Single-ring Sample}
}
}

\vspace{2mm}

The direction of neutrino for FC single-ring sample is 
simply assumed to be the same as the reconstructed direction of muon.
Zenith angle of neutrino is reconstructed as follows:
\[
\hspace{2cm}\cos\Theta^{rec}_{\nu}=\cos\Theta_{\mu} \hspace{2cm}(8.17) 
\]
,where $\cos\Theta^{rec}_{\nu}$ and $\cos\Theta_{\mu}$ are 
cosine of the reconstructed zenith angle of neutrino and muon,
respectively.}" 
\end{quote}

[C] Jung, Kajita {\it et al.} \cite{Jung} state
\footnote{see page 453 of the paper concerned.}:
\begin{quote}
"{\it At neutrino energies of more than a few hundred MeV, the 
direction of the reconstructed lepton approximately represents 
the direction of the original neutrino. Hence, for detectors near 
the surface of the Earth, the neutrino flight distance is a function of
the zenith direction of the lepton. Any effects, 
such as neutrino oscillations, that are a function of the neutrino
 flight distance will be manifest in the 
lepton zenith angle distributions.}" 
\end{quote}

 Hereafter, we call the fundamental assumption by 
the Super-Kamiokande Experiment simply as 
{\it the SK assumption on the direction}. 

Among various types of the neutrino events analyzed by Super-Kamiokande,
 the most important events are single ring muon(electron) events 
which are generated in the detector and terminate in it,
 because these events give more essential physical quantities for clear 
interpretation on neutrino oscillation,
namely, the kinds of the neutrino, the transferred 
energies and the directions of the produced leptons.
 These single ring muon events are generated from the quasi elastic
 scattering (QEL). 
If the existence of neutrino oscillation is verified surely
in the analyses
of single ring muon events among {\it Fully Contained Events}
under {\it the SK assumption on the direction},
then we can expect the corroboration
of the oscillation in the analyses of other types of the
interaction with less accuracy.
Therefore,  {\it the SK assumption on the direction} should be 
carefully examined in the analysis of single ring events due to 
QEL among {\it Fully Contained Events}. 

Our paper is organized as follows.
 In section~2, we treat the differential cross section for QEL in the
stochastic manner as exactly as possible and obtain the zenith angle 
distributions of the emitted leptons, $\cos\theta_\mu$,
 for given neutrinos with definite zenith angles, taking account of 
the azimuthal angles of the emitted leptons in QEL.
 As a result of it, we show that 
{\it the SK assumption on the direction} does not hold any more for 
the incident neutrinos with smaller energies. 
In section 3, we examine 
{\it the SK assumption on the direction} in the light 
of  $L_{\nu}$ and $L_{\mu}$
and reach the same conclusion obtained in the section 2, as it must be. 
In section 4, we examine all possible choices of combination of 
$L$ and $E$, namely, $L_{\nu}/E_{\nu}$, $L_{\nu}/E_{\mu}$,
$L_{\mu}/E_{\nu}$, $L_{\mu}/E_{\mu}$,
 in $L/E$ analysis and show that the maximum oscillation can be realized 
in the combination of $L_{\nu}/E_{\nu}$, as it must be.
 In section 5, we compare our results obtained from the numerical 
experiments with real observation obtained by 
Super-Kamiokande Collaboration.
 In section 6, we conclude that SK cannot observe the maximum oscillation in their $L_{\mu}/E_{\nu}$ analysis.  

Here, we designate the minimum extrema in $L/E$ distribution for 
neutrino oscillation as the maximum oscillation by the terminology  
already utilized in the Super-Kamiokande Collaboration \cite{Ashie1}.

\section{Single Ring Events among Fully Contained 
Events which are Produced by Quasi Elastic Scsattering.
}

\subsection{
Differential cross section of quasi elastic scattering and influence 
over various quantities concerned
}
 As stated in Introduction, the finding of the observation of 
the maximum oscillation in the $L/E$ analysis is the ultimate 
verification of the finding of the neutrino 
oscillation by Super-Kamiokande. 
 For the examination of the Super-Kamiokande's assertion, we analyze 
the $L/E$ distribution of the single ring events among 
{\it Fully Contained Events}.

 In order to examine the validity of {\it the SK assumption on 
the direction},  we consider the single ring events due to
the following quasi elastic scattering(QEL):

   \begin{eqnarray}
         \nu_e + n \longrightarrow p + e^-  \nonumber\\
         \nu_{\mu} + n \longrightarrow p + \mu^- \nonumber\\
         \bar{\nu}_e + p \longrightarrow n + e^+ \\
         \bar{\nu}_{\mu}+ p \longrightarrow n + \mu^+ \nonumber
         ,\label{eqn:1}
   \end{eqnarray}
The differential cross section for QEL is given as follows \cite{r4}.\\
    \begin{eqnarray}
         \frac{ {\rm d}\sigma_{\ell(\bar{\ell})}(E_{\nu(\bar{\nu})}) }{{\rm d}Q^2} = 
         \frac{G_F^2{\rm cos}^2 \theta_C}{8\pi E_{\nu(\bar{\nu})}^2}
         \Biggl\{ A(Q^2) \pm B(Q^2) \biggl[ \frac{s-u}{M^2} \biggr]
         + \nonumber \\ 
C(Q^2) \biggl[ \frac{s-u}{M^2} \biggr]^2 \Biggr\},
         \label{eqn:2}
    \end{eqnarray}

\noindent where
    \begin{eqnarray*}
      A(Q^2) &=& \frac{Q^2}{4}\Biggl[ f^2_1\biggl( \frac{Q^2}{M^2}-4 \biggr)+ f_1f_2 \frac{4Q^2}{M^2} \\
 &&+  f_2^2\biggl( \frac{Q^2}{M^2} -\frac{Q^4}{4M^4} \biggr) + g_1^2\biggl( 4+\frac{Q^2}{M^2} \biggl) \Biggr], \\
      B(Q^2) &=& (f_1+f_2)g_1Q^2, \\
      C(Q^2) &=& \frac{M^2}{4}\biggl( f^2_1+f^2_2\frac{Q^2}{4M^2}+g_1^2 \biggr).
    \end{eqnarray*}

\noindent The signs $+$ and $-$ refer to $\nu_{\mu(e)}$ and $\bar{\nu}_{\mu(e)}$ for charged current (c.c.) interactions, respectively.  The $Q^2$ denotes the four momentum transfer between the incident neutrino and the charged lepton. Details of other symbols are given in \cite{r4}.

The relation among $Q^2$, $E_{\nu(\bar{\nu})}$, the energy of the incident neutrino, $E_{\ell}$, the energy of the emitted charged lepton (muon or electron or their anti-particles) and $\theta_{\rm s}$, the scattering angle of the emitted lepton, is given as
      \begin{equation}
         Q^2 = 2E_{\nu(\bar{\nu})}E_{\ell}(1-{\rm cos}\theta_{\rm s}).
                  \label{eqn:3}
      \end{equation}

\noindent Also, the energy of the emitted lepton is given by
      \begin{equation}
         E_{\ell} = E_{\nu(\bar{\nu})} - \frac{Q^2}{2M}.
\label{eqn:4}
      \end{equation}

Now, let us examine the magnitude of the scattering angle of the emitted 
lepton in a quantitative way, as this plays a decisive role in determining 
the accuracy of the direction of the incident neutrino,
which is directly related to the reliability of the zenith angle 
distribution of 
single ring muon (electron) events in the Super-Kamiokande Experiment.

By using Eqs. (\ref{eqn:2}) to (\ref{eqn:4}), we obtain the distribution 
function for the scattering angle of the emitted leptons and the related 
quantities by a Monte Carlo method. The procedure for determining the 
scattering angle for a given energy of the incident neutrino is described 
in Appendix A.  Figure~\ref{figH001} shows this relation for muon, from 
which we can easily understand that the scattering angle $\theta_{\rm s}$ 
of the emitted lepton ( muon here ) cannot be neglected.  For a 
quantitative examination of the scattering angle, we construct the 
distribution function for ${\theta_{\rm s}}$ of the emitted lepton from 
Eqs. (\ref{eqn:2}) to (\ref{eqn:4}) by using the Monte Carlo method.

Figure~\ref{figH002} gives the distribution function for $\theta_{\rm s}$ of 
the muon produced in the muon neutrino interaction. It can be seen that 
the muons produced from lower energy neutrinos are scattered over wider 
angles and that a considerable part of them are scattered even in backward 
directions. 
Similar results are obtained for anti-muon neutrinos, electron neutrinos 
and anti-electron neutrinos.
\begin{figure}
\begin{center}
\vspace{-1cm}
\hspace*{-1cm}
\rotatebox{90}{%
\resizebox{0.35\textwidth}{!}{\includegraphics{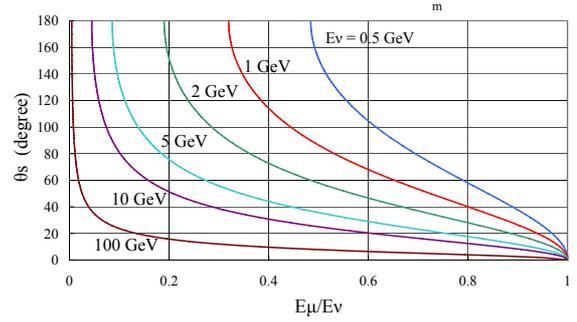}}}

\vspace{-1cm}
\caption{Relation between the energy of the muon and its 
scattering angle for different incident muon neutrino energies,
 0.5, 1, 2, 5, 10 and 100~GeV.}
\label{figH001}
\end{center}
\end{figure} 
\begin{figure}
\begin{center}
\vspace{-1.5cm}
\rotatebox{90}{%
\resizebox{0.45\textwidth}{!}{\includegraphics{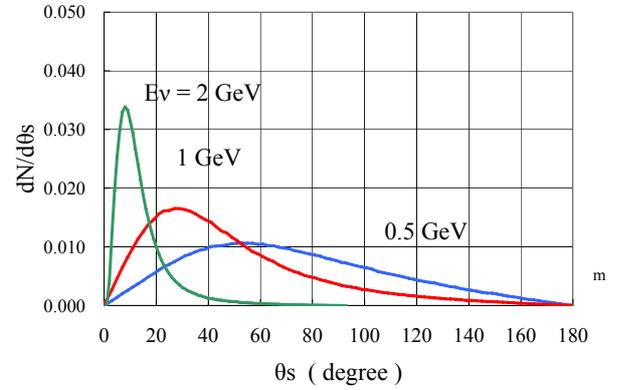}}}
\vspace{-1cm}
\caption{Distribution functions for the scattering angle of 
the muon for muon-neutrino with incident energies, 0.5 , 1.0 and 
2~GeV. Each curve is obtained by the Monte Carlo 
method (one million sampling per each curve). }
\label{figH002}
\end{center}
\end{figure} 

Also, in a similar manner, we obtain not only the distribution function 
for the scattering angle of the charged leptons, but also their average 
values $<\theta_{\rm s}>$ and their standard deviations $\sigma_{\rm s}$. 
Table~1 shows them  for muon neutrinos, anti-muon neutrinos, electron 
neutrinos and anti-electron neutrinos.  
From Table~1, it seems to be clear that the scattering angles 
could not be neglected, taking account of the fact that
 the frequency of the neutrino events with smaller energies is far 
larger than that of the neutrino events with larger energies 
due to high steep of the neutrino energy spectrum.  
 However,   
Super-Kamiokande Collaboration assume that the direction of the neutrino 
is approximately the same as that of the emitted lepton even for the 
neutrino events with smaller energies,
as cited in {\it the three
passages} mentioned above \cite{Kajita1},
\cite{Ishitsuka},\cite{Jung}. 
However, it has never been verified by Super-Kamiokande Collaboration.

\begin{table*}
\caption{\label{tab:table1} The average values $<\theta_{\rm s}>$ for 
scattering angle of the emitted charged leptons and their standard 
deviations $\sigma_{\rm s}$ for various primary neutrino energies 
$E_{\nu(\bar{\nu})}$.}
\vspace{5mm}
\begin{center}
\begin{tabular}{|c|c|c|c|c|c|}
\hline
&&&&&\\
$E_{\nu(\bar{\nu})}$ (GeV)&angle&$\nu_{\mu(\bar{\mu})}$&
$\bar{\nu}_{\mu(\bar{\mu})}$&$\nu_e$&$\bar{\nu_e}$ \\
&(degree)&&&&\\
\hline
0.2&$<\theta_\mathrm{s}>$&~~ 89.86 ~~&~~ 67.29 ~~&~~ 89.74 ~~&~~ 67.47 ~~\\
\cline{2-6}
   & $\sigma_\mathrm{s}$ & 38.63 & 36.39 & 38.65 & 36.45 \\
\hline
0.5&$<\theta_\mathrm{s}>$& 72.17 & 50.71 & 72.12 & 50.78 \\
\cline{2-6}
   & $\sigma_\mathrm{s}$ & 37.08 & 32.79 & 37.08 & 32.82 \\
\hline
1  &$<\theta_\mathrm{s}>$& 48.44 & 36.00 & 48.42 & 36.01 \\
\cline{2-6}
   & $\sigma_\mathrm{s}$ & 32.07 & 27.05 & 32.06 & 27.05 \\
\hline
2  &$<\theta_\mathrm{s}>$& 25.84 & 20.20 & 25.84 & 20.20 \\
\cline{2-6}
   & $\sigma_\mathrm{s}$ & 21.40 & 17.04 & 21.40 & 17.04 \\
\hline
5  &$<\theta_\mathrm{s}>$&  8.84 &  7.87 &  8.84 &  7.87 \\
\cline{2-6}
   & $\sigma_\mathrm{s}$ &  8.01 &  7.33 &  8.01 &  7.33 \\
\hline
10 &$<\theta_\mathrm{s}>$&  4.14 &  3.82 &  4.14 &  3.82 \\
\cline{2-6}
   & $\sigma_\mathrm{s}$ &  3.71 &  3.22 &  3.71 &  3.22 \\
\hline
100&$<\theta_\mathrm{s}>$&  0.38 &  0.39 &  0.38 &  0.39 \\
\cline{2-6}
   & $\sigma_\mathrm{s}$ &  0.23 &  0.24 &  0.23 &  0.24 \\
\hline
\end{tabular}
\end{center}
\label{tab:1}
\end{table*}

%
\subsection{Influence of azimuthal angle in QEL
 over the zenith angle of single ring events}

In the present subsection, we examine the effect of the azimuthal angles,
$\phi$, of the emitted leptons over their own zenith angles,
$\theta_{\mu(\bar{\mu}))}$, for given zenith angles 
of the incident neutrinos, $\theta_{\nu(\bar{\nu}))}$ in QEL,
 which was not be considered in the detector simulation
carried by Super-Kamiokande Collaboration
\footnote{Throughout this paper, we measure the zenith angles of the 
emitted leptons from the upward vertical direction of the incident 
neutrino. Consequently, notice that the sign of our direction is opposite 
to that of the Super-Kamiokande Experiment 
( our $\cos\theta_{\nu(\bar{\nu})}$ = - 
$\cos\theta_{\nu(\bar{\nu})}$ in SK)}.
The influence of this effect over the zenith angle cannot be neglected
particularly in horizontal-like neutrino events.

\begin{figure}
\begin{center}
\vspace{-0.5cm}
\resizebox{0.5\textwidth}{!}{%
  \includegraphics{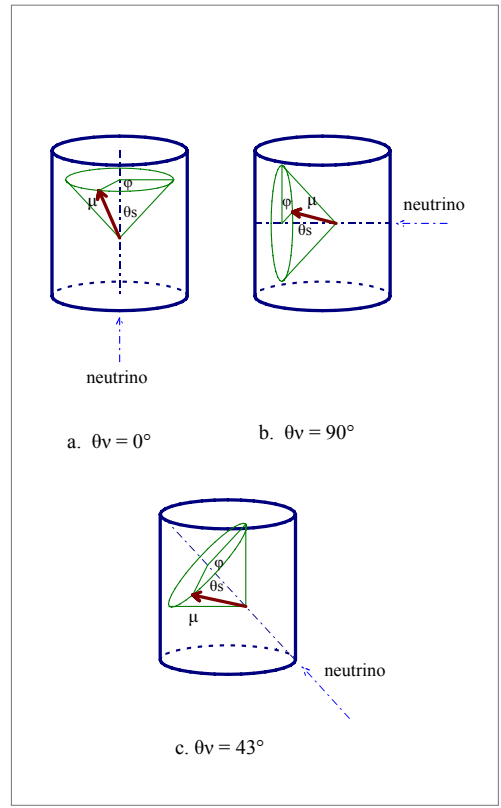}
  }
\end{center}
\caption{Schematic view of the zenith angles of the charged
 muons for different zenith angles of the incident neutrinos, focusing on
  their azimuthal angles.}
\label{figH003}
\end{figure}

For three typical cases (vertical, horizontal and diagonal), 
Figure~\ref{figH003} gives 
a schematic representation of the relationship between, 
$\theta_{\nu(\bar{\nu})}$, the zenith angle of the incident neutrino, and (
$\theta_{\rm s}$, $\phi$), a pair of scattering angle of the emitted lepton 
and its azimuthal angle.  

From Figure~\ref{figH003}-a, it can been seen that the zenith angle 
$\theta_{\mu(\bar{\mu})}$ of the emitted lepton is not influenced by its 
$\phi$ in the vertical incidence of the neutrinos 
$(\theta_{\nu(\bar{\nu})}=0^{\rm o})$, as it must be. From 
Figure~\ref{figH003}-b, 
however, it is obvious that the influence of $\phi$ of the emitted leptons 
on their own zenith angle is the strongest in the case of horizontal 
incidence of the neutrino $(\theta_{\nu(\bar{\nu})}=90^{\rm o})$. Namely, 
one half of the emitted leptons are recognized as upward going, while the 
other half is classified as downward going ones. The diagonal case ( 
$\theta_{\nu(\bar{\nu})}=43^{\rm o}$) is intermediate between the vertical 
and the horizontal. In the following, we examine the cases for vertical, 
horizontal and diagonal incidence of the neutrinos with different
energies, 
say, $E_{\nu(\bar{\nu})}=0.5$ GeV, $E_{\nu(\bar{\nu})}=1$ GeV and 
$E_{\nu(\bar{\nu})}=5$ GeV,
as the typical cases. 

\subsection{Dependence of the spreads of the zenith angle for the emitted 
leptons on the energies 
of the emitted leptons for different incident directions 
of the neutrinos with different 
energies}
\begin{figure*}
\hspace{2.5cm}(a)
\hspace{5.5cm}(b)
\hspace{5.5cm}(c)
\vspace{-0.3cm}
\begin{center}
\resizebox{\textwidth}{!}{%
  \includegraphics{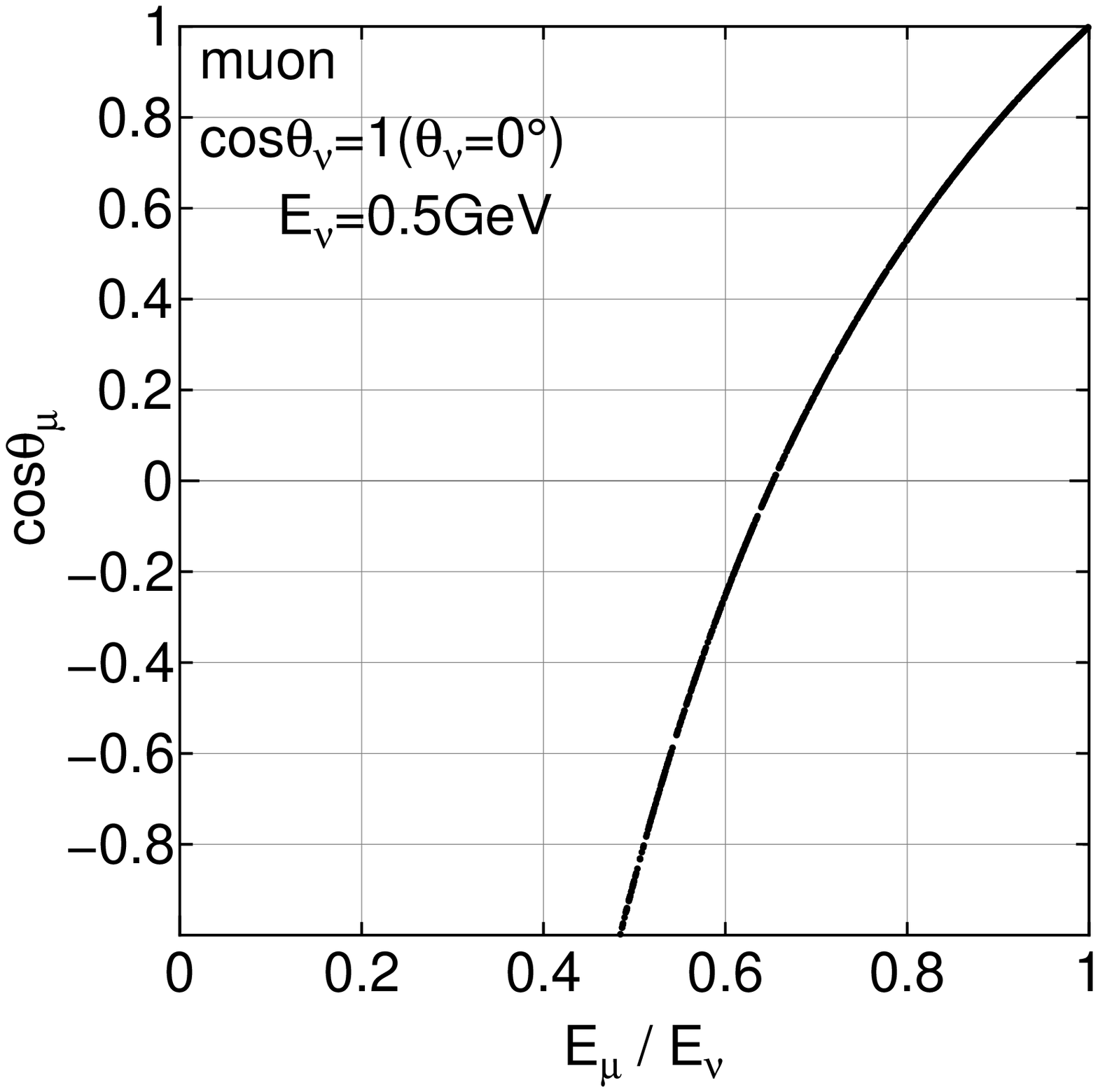}\hspace{1cm}
  \includegraphics{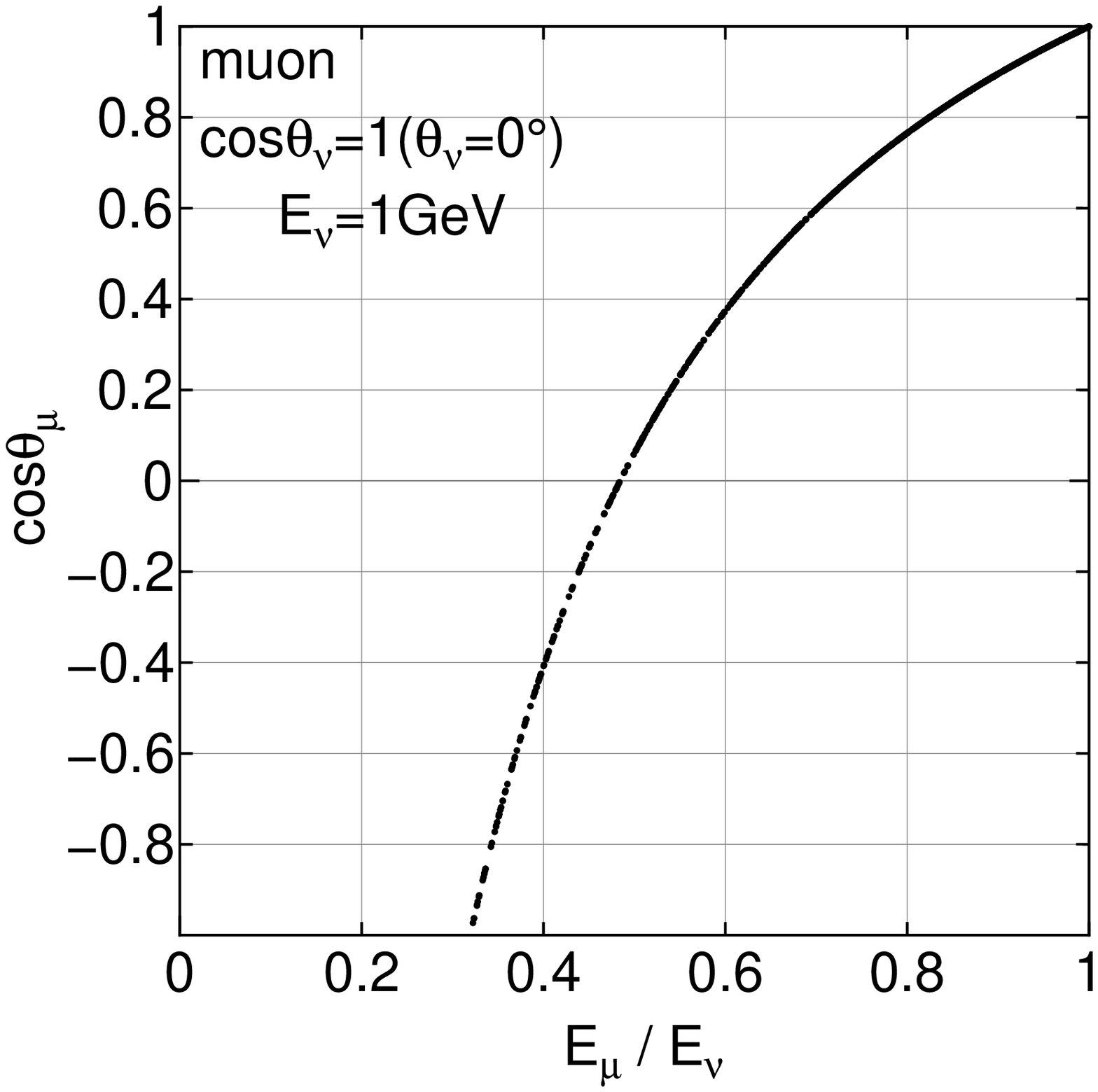}\hspace{1cm}
  \includegraphics{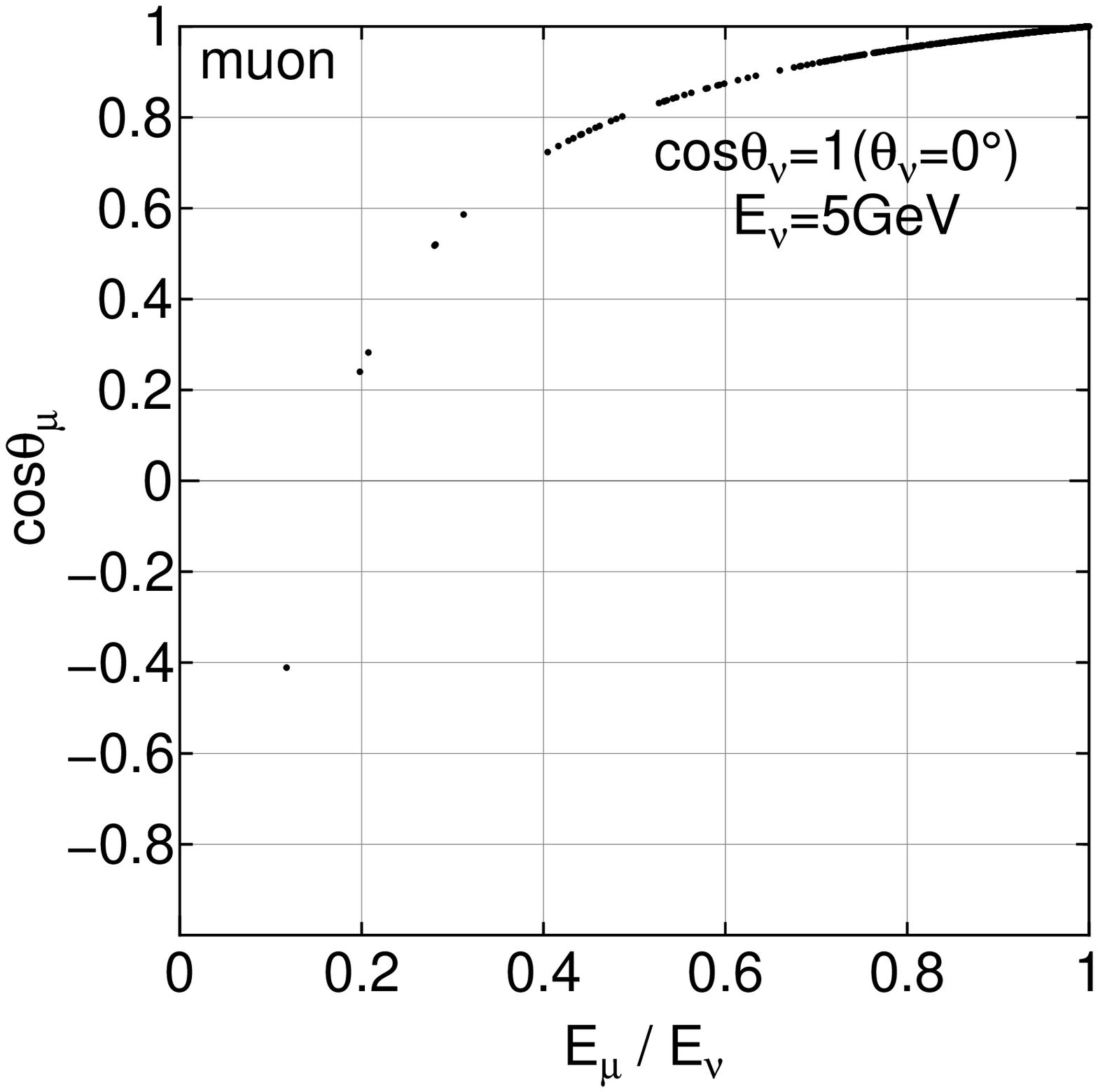}
}
\caption{
\label{figH004}
The scatter plots between the fractional energies of the produced muons 
and their zenith angles 
for vertically incident muon neutrinos with 0.5~GeV, 1~GeV and 5~GeV, 
respectively.
 The sampling number is 1000 for each case.
}
\end{center}
\vspace{0.5cm}
\hspace{2.5cm}(a)
\hspace{5.5cm}(b)
\hspace{5.5cm}(c)
\vspace{-0.3cm}
\begin{center}
\resizebox{\textwidth}{!}{%
  \includegraphics{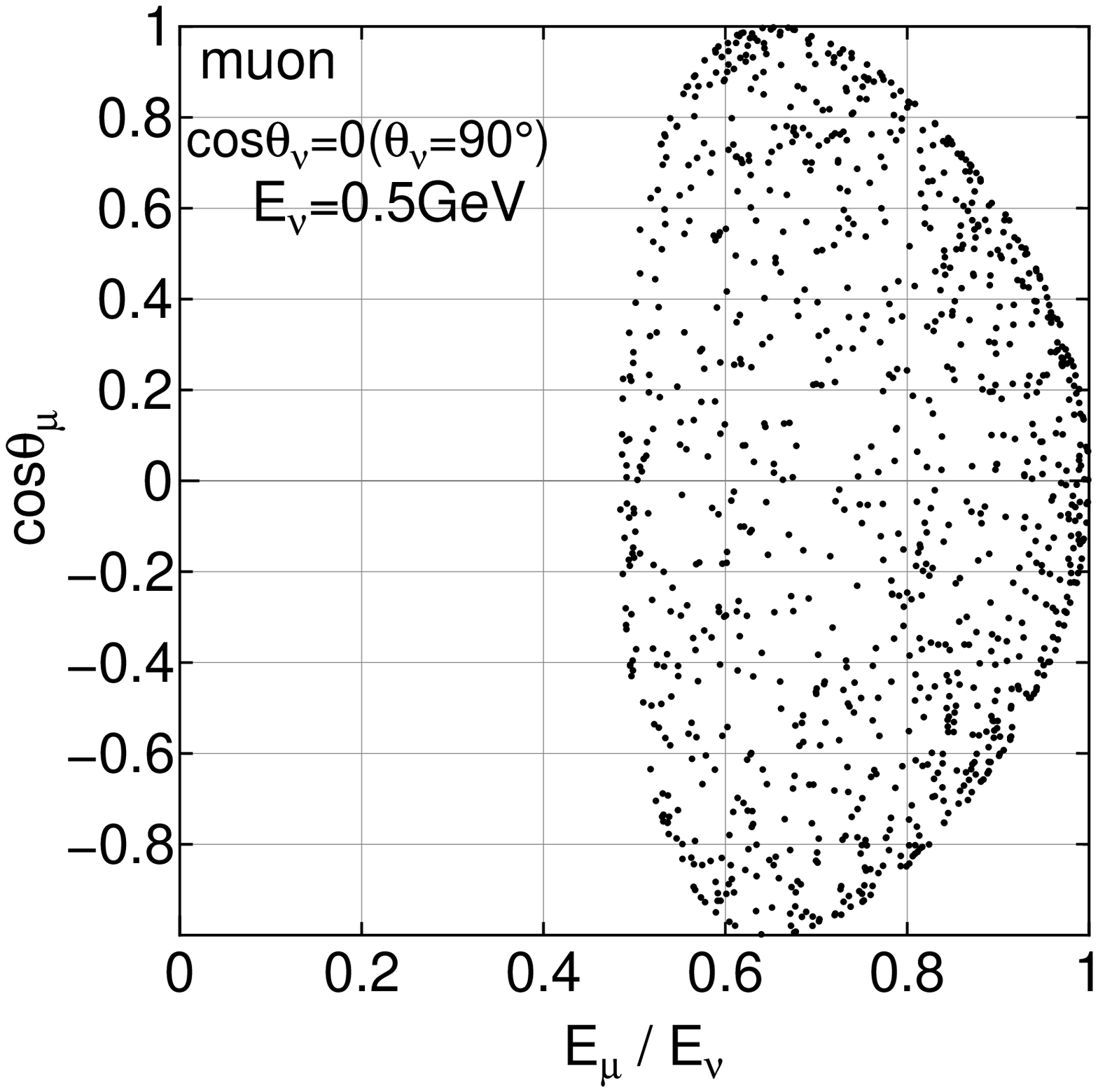}\hspace{1cm}
  \includegraphics{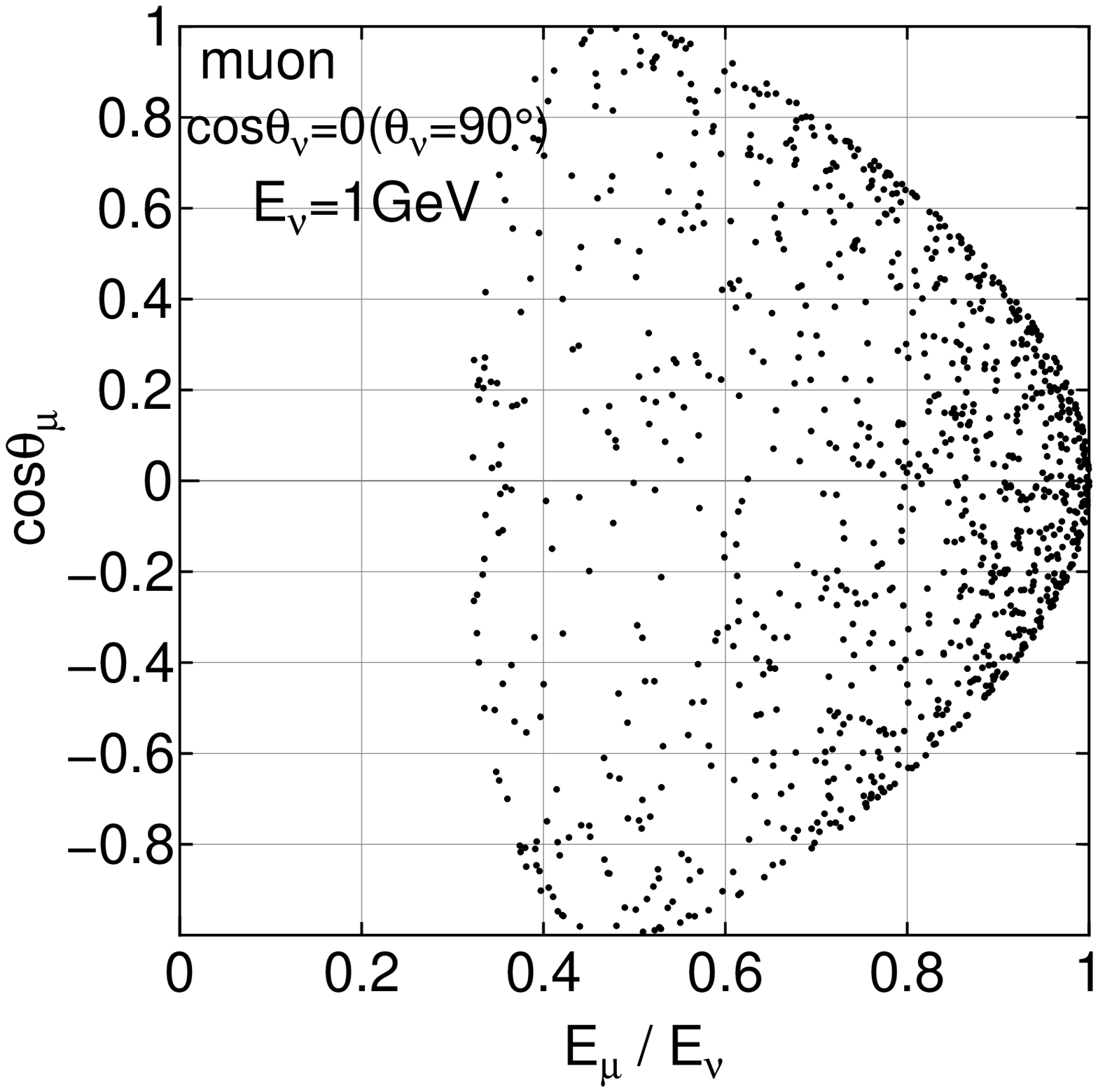}\hspace{1cm}
  \includegraphics{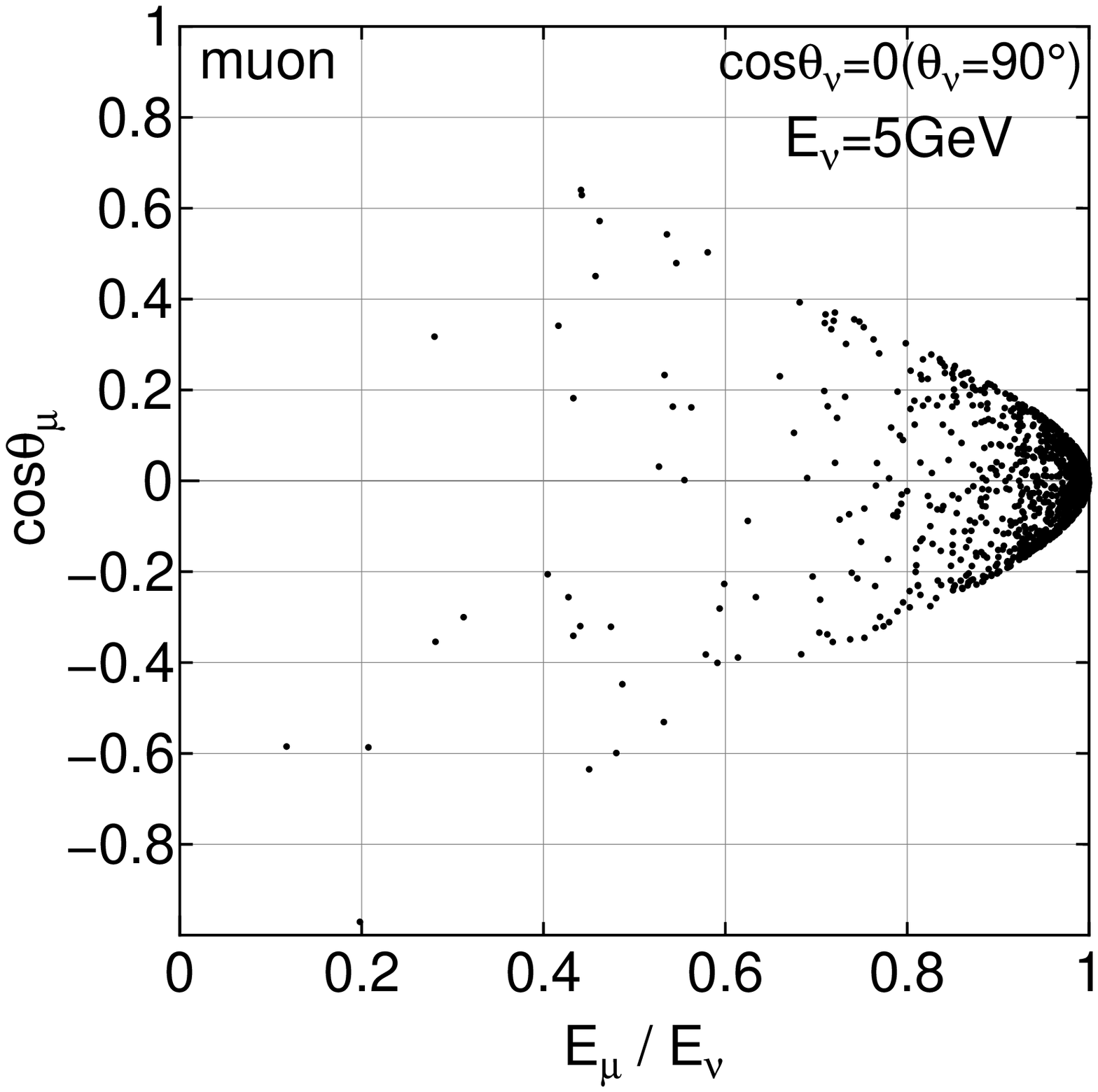}
}
\caption{
\label{figH005} 
The scatter plots between the fractional energies of the produced muons 
and their zenith angles 
for horizontally incident muon neutrinos with 0.5~GeV, 1~GeV and 5~GeV, 
respectively.
 The sampling number is 1000 for each case.
}
\end{center}
\vspace{0.5cm}
\hspace{2.5cm}(a)
\hspace{5.5cm}(b)
\hspace{5.5cm}(c)
\vspace{-0.3cm}
\begin{center}
\resizebox{\textwidth}{!}{%
  \includegraphics{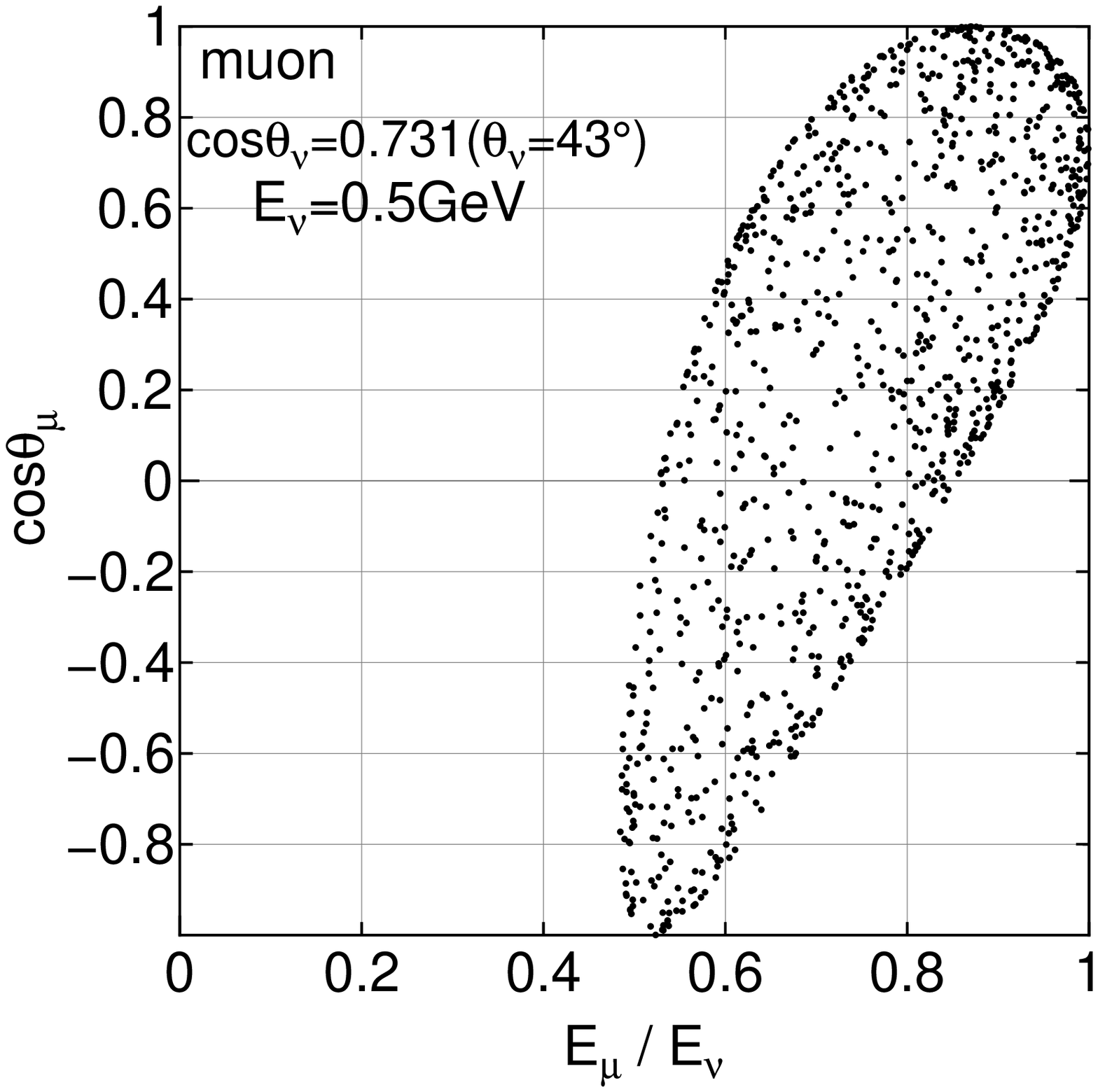}\hspace{1cm}
  \includegraphics{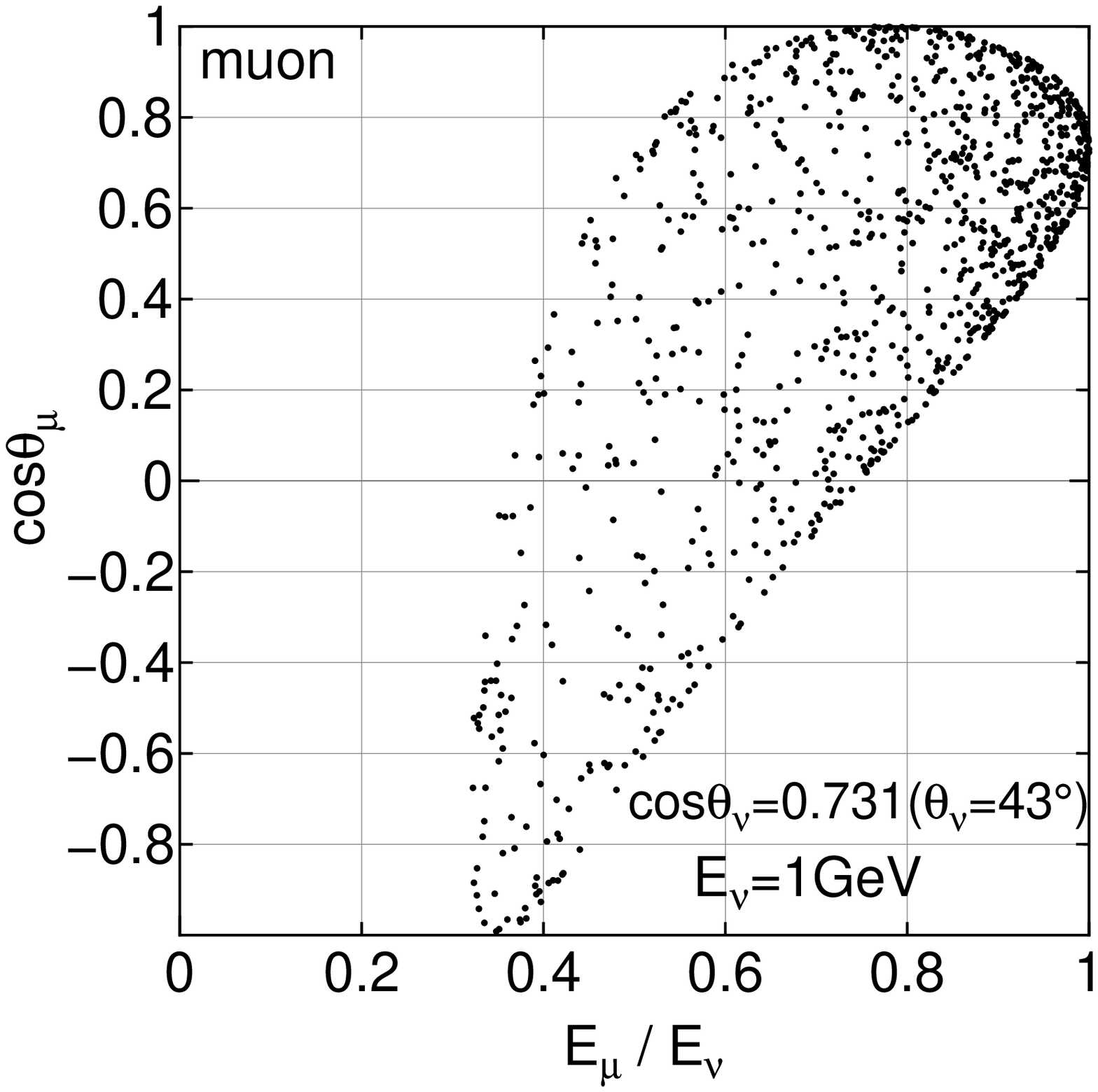}\hspace{1cm}
  \includegraphics{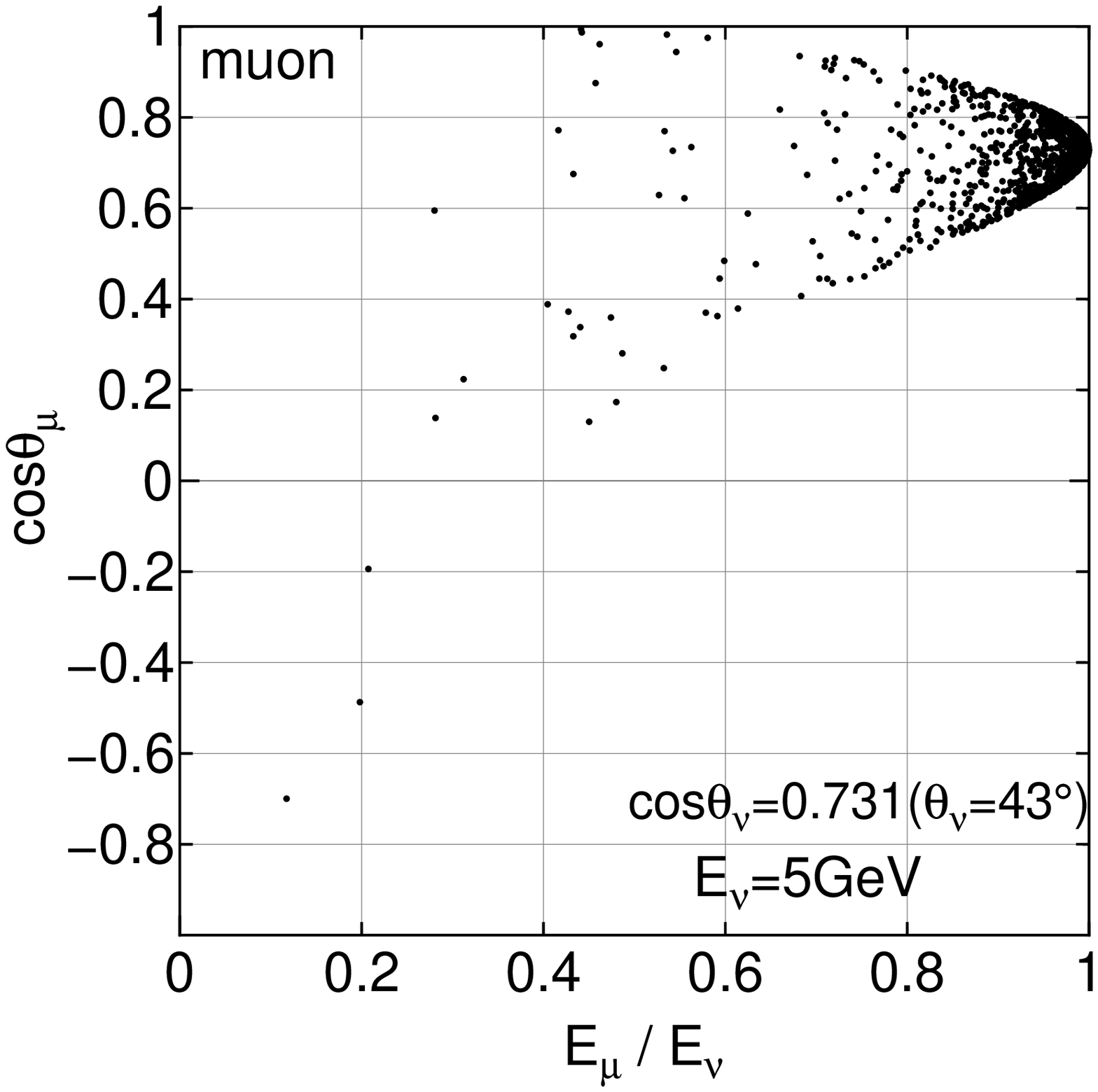}
  }
\caption{
\label{figH006} 
The scatter plots between the fractional energies of the produced muons 
and their zenith angles 
for diagonally incident muon neutrinos with 0.5~GeV, 1~GeV and 5~GeV, 
respectively.
 The sampling number is 1000 for each case.
}
\end{center}
\end{figure*}

  The detailed procedure for the Monte Carlo simulation is described in 
 Appendix A. 
 We give  the scatter plots between the fractional energies of the emitted 
muons and their zenith angle
 for a definite zenith angle of the incident neutrino with different 
energies in Figures~\ref{figH004} to \ref{figH006}.
 In Figure~\ref{figH004}, we give 
the scatter plots for vertically incident neutrinos with different 
energies 
0.5, 1 and 5~GeV. In this case, the relations between the emitted 
energies of the muons and their zenith angles are unique, which comes 
from the definition of the zenith angle of the emitted lepton. However, 
the densities (frequencies of the event number) along each curves are 
different 
in position to position and depend on the energies of the incident 
neutrinos. Generally speaking, densities along the curves become greater 
toward  $\cos\theta_{\mu(\bar{\mu})}= 1$. In this case, 
$\cos\theta_{\mu(\bar{\mu})}$ is never influenced by the azimuthal angel 
in the scattering by the definition
\footnote{The zenith angles of the particles concerned are measured from 
the vertical direction.}.
 
On the contrast, it is shown in Figure~\ref{figH005} 
that the horizontally incident 
neutrinos give the widest zenith angle distributions for the 
emitted muons with the same fractional energies
 due to the effect of the azimuthal angles.
 The lower the energies of the incident neutrinos are, the  
wider the spreads of the scattering angles of emitted muons
$\theta_{\mu}$ become,
which leads to wider zenith angle distributions for the emitted muons.
  As easily understood from Figure~\ref{figH006}, 
the diagonally incident neutrinos give the intermediate zenith angle 
distributions for the emitted  muons between those for vertically 
incident neutrinos and those for horizontally incident neutrinos.  

\begin{figure*}
\hspace{2.5cm}(a)
\hspace{5.5cm}(b)
\hspace{5.5cm}(c)
\vspace{-0.3cm}
\begin{center}
\resizebox{\textwidth}{!}{%
  \includegraphics{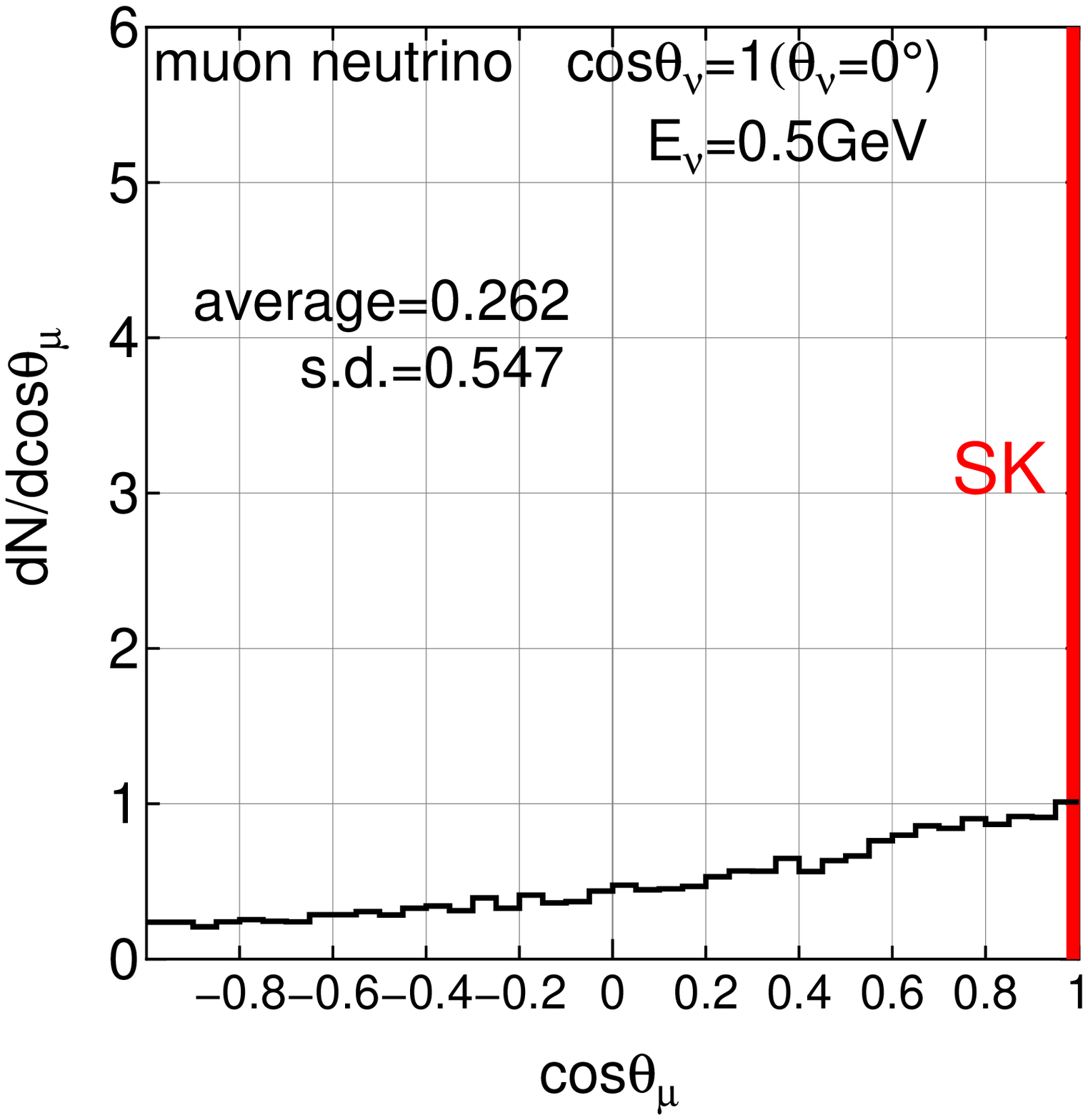}\hspace{1cm}
  \includegraphics{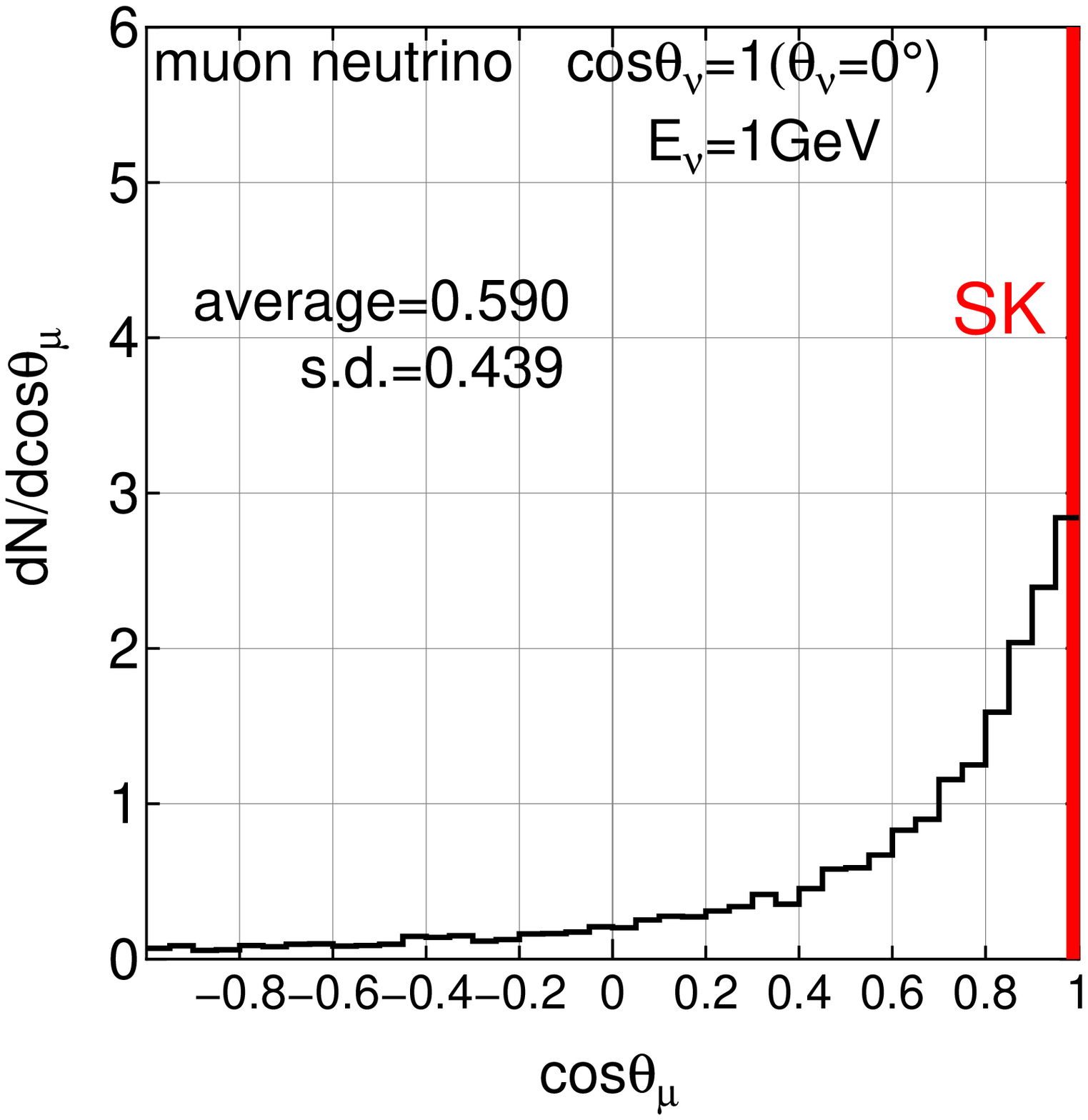}\hspace{1cm}
  \includegraphics{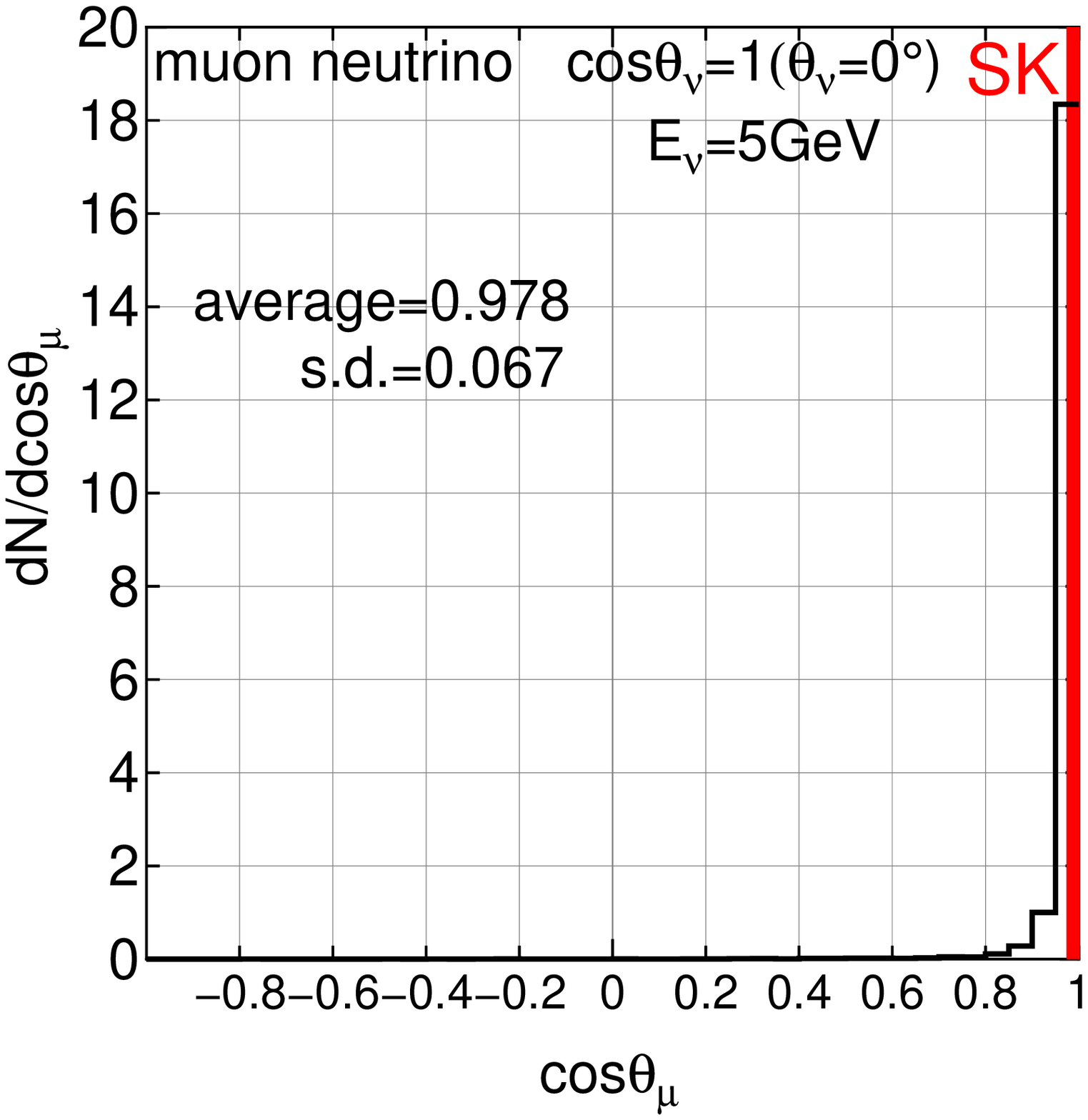}
  }
\caption{
\label{figH007} 
Zenith angle distribution of the muon for the vertically incident muon 
neutrino with 0.5~GeV, 1~GeV and 5~GeV, respectively. The sampling 
number is 10000 for each case.
SK stands for the corresponding ones under the SK assumption.
}
\end{center}
\vspace{0.5cm}

\hspace{2.5cm}(a)
\hspace{5.5cm}(b)
\hspace{5.5cm}(c)
\vspace{-0.3cm}
\begin{center}
\resizebox{\textwidth}{!}{%
 \includegraphics{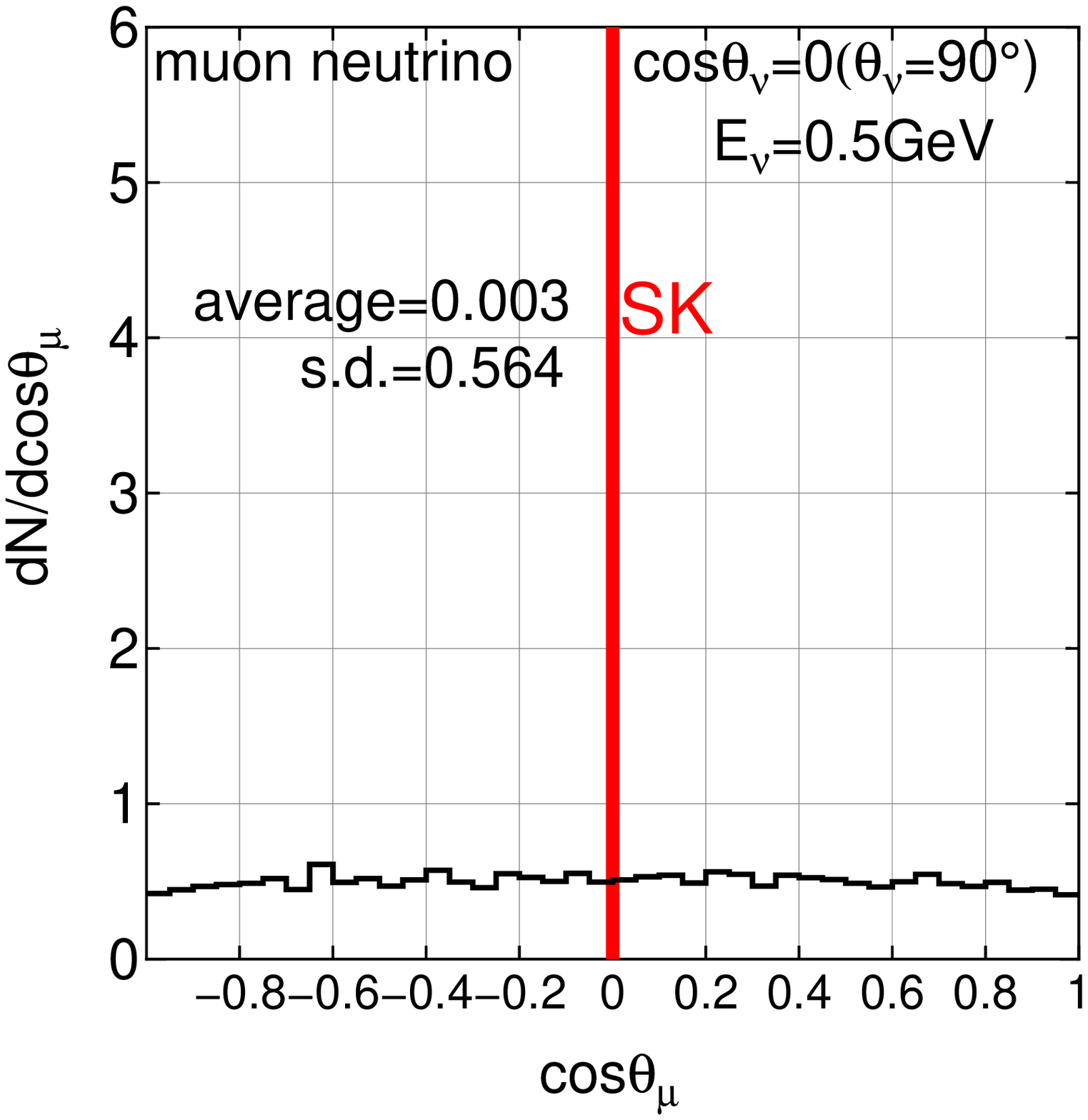}\hspace{1cm}
  \includegraphics{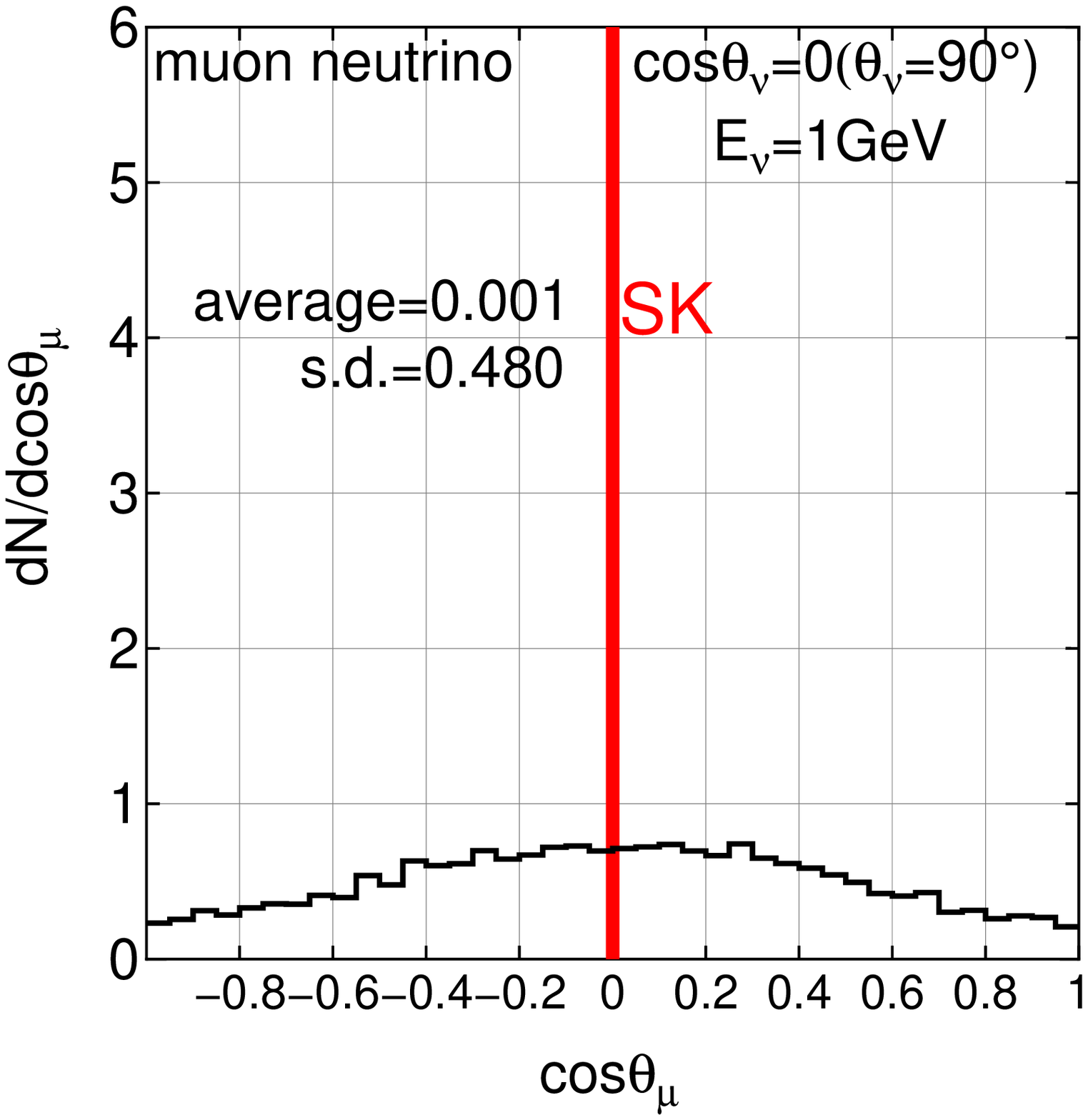}\hspace{1cm}
  \includegraphics{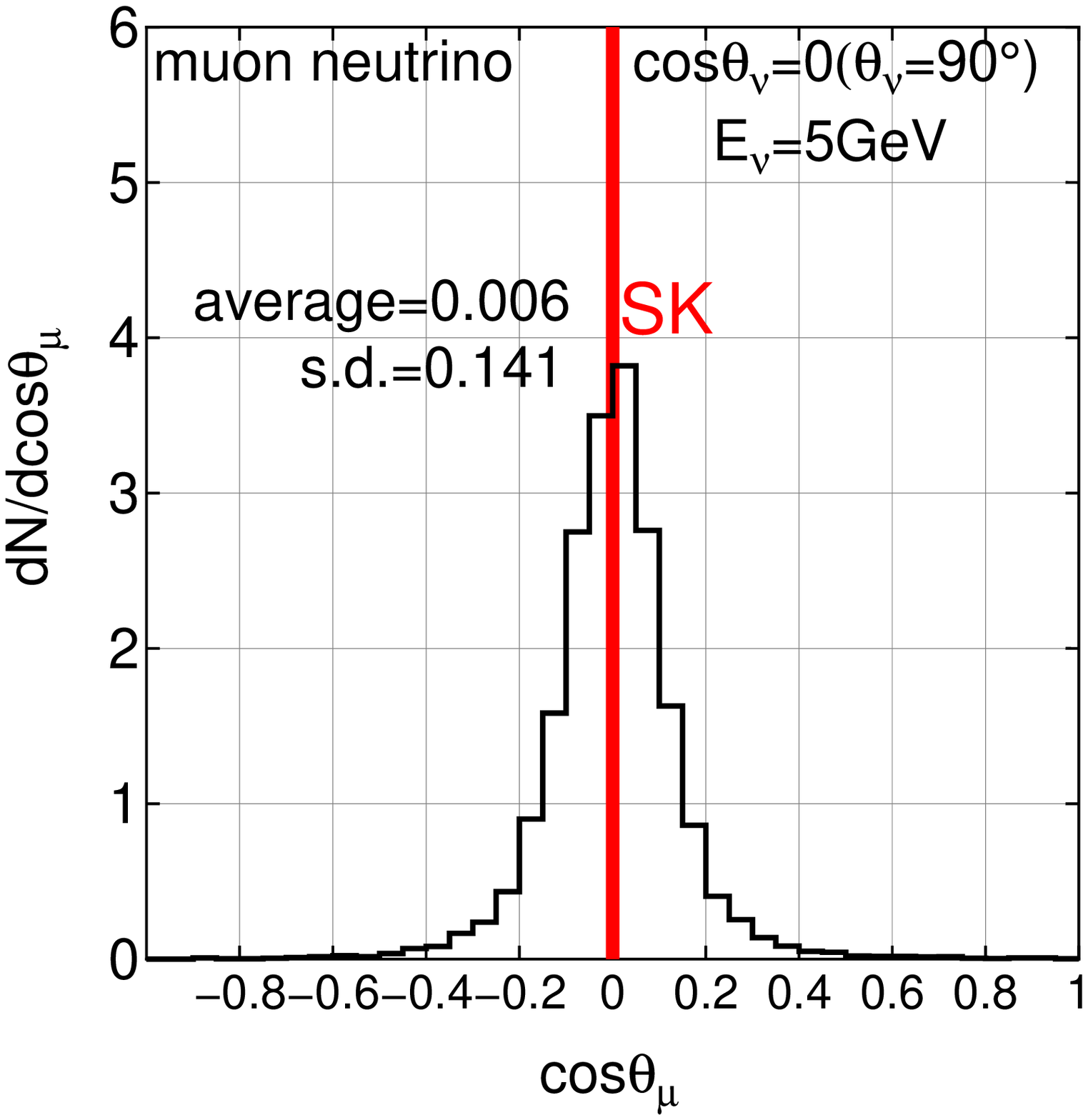}
  }
\caption{
\label{figH008} 
Zenith angle distribution of the muon for the horizontally incident muon 
neutrino with 0.5~GeV, 1~GeV and 5~GeV, respectively. The sampling number 
is 10000 for each case.
SK stands for the corresponding ones under the SK assumption.
}
\end{center}
\vspace{0.5cm}
\hspace{2.5cm}(a)
\hspace{5.5cm}(b)
\hspace{5.5cm}(c)
\vspace{-0.3cm}
\begin{center}
\resizebox{\textwidth}{!}{%
  \includegraphics{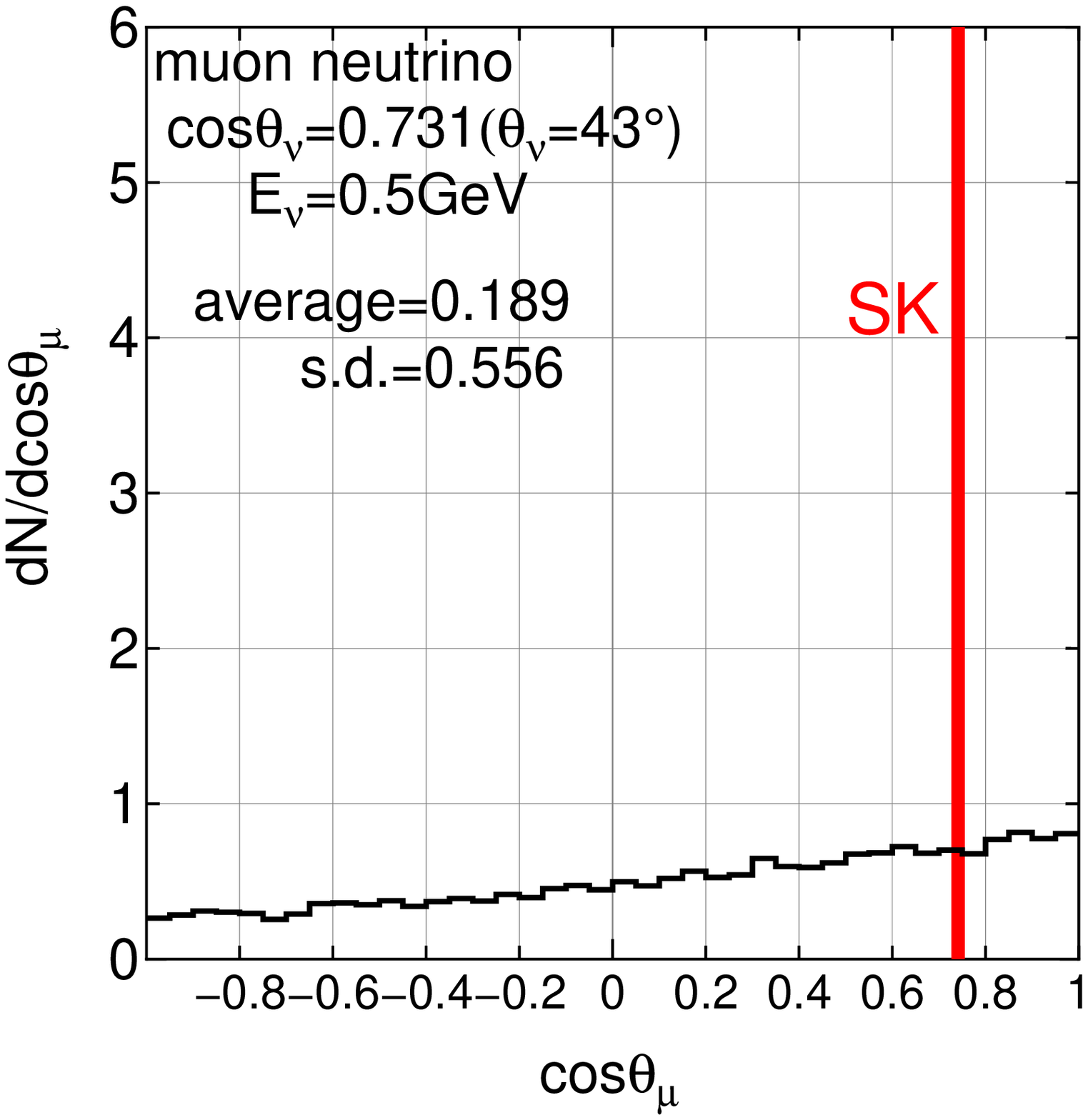}\hspace{1cm}
  \includegraphics{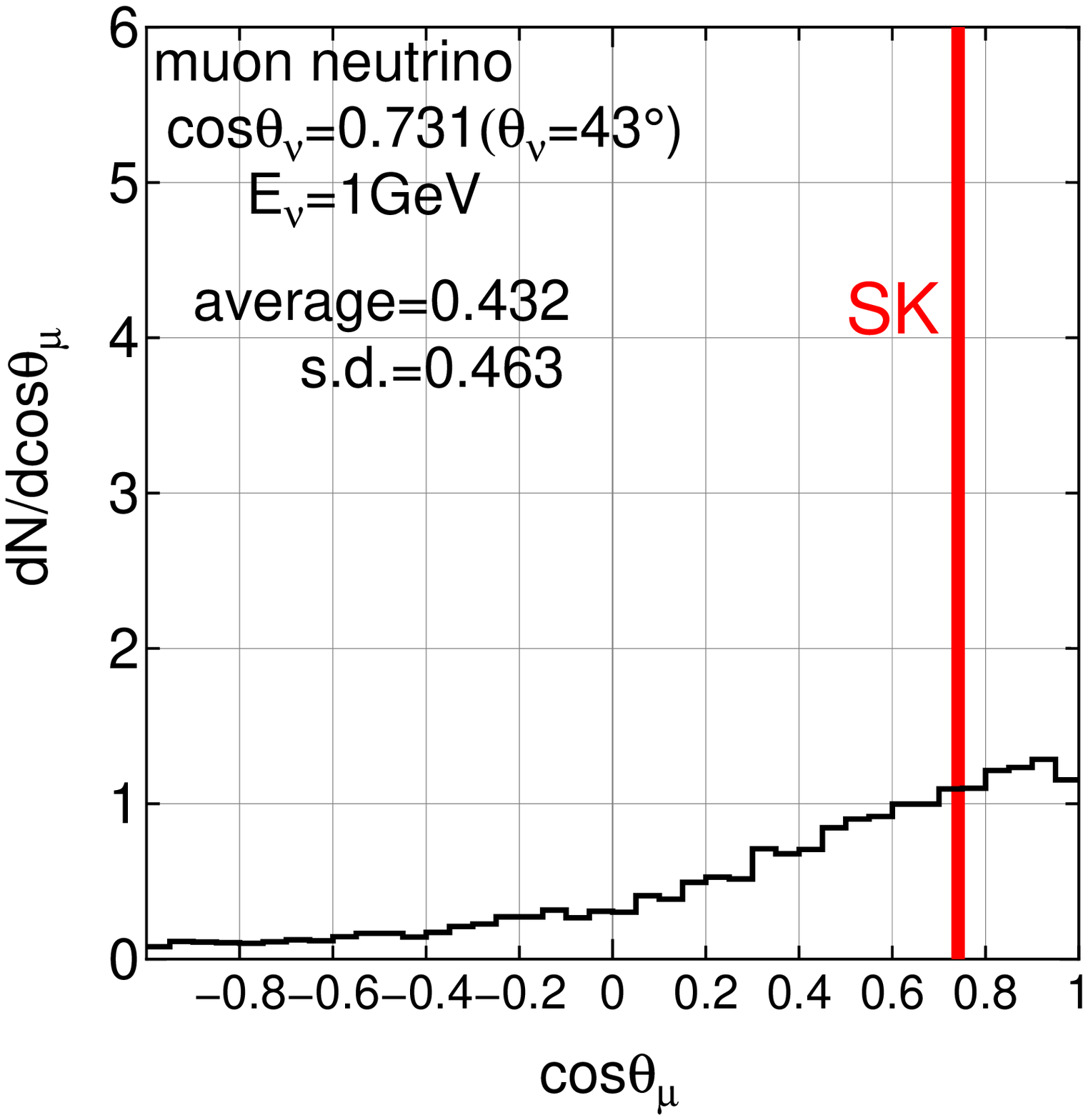}\hspace{1cm}
  \includegraphics{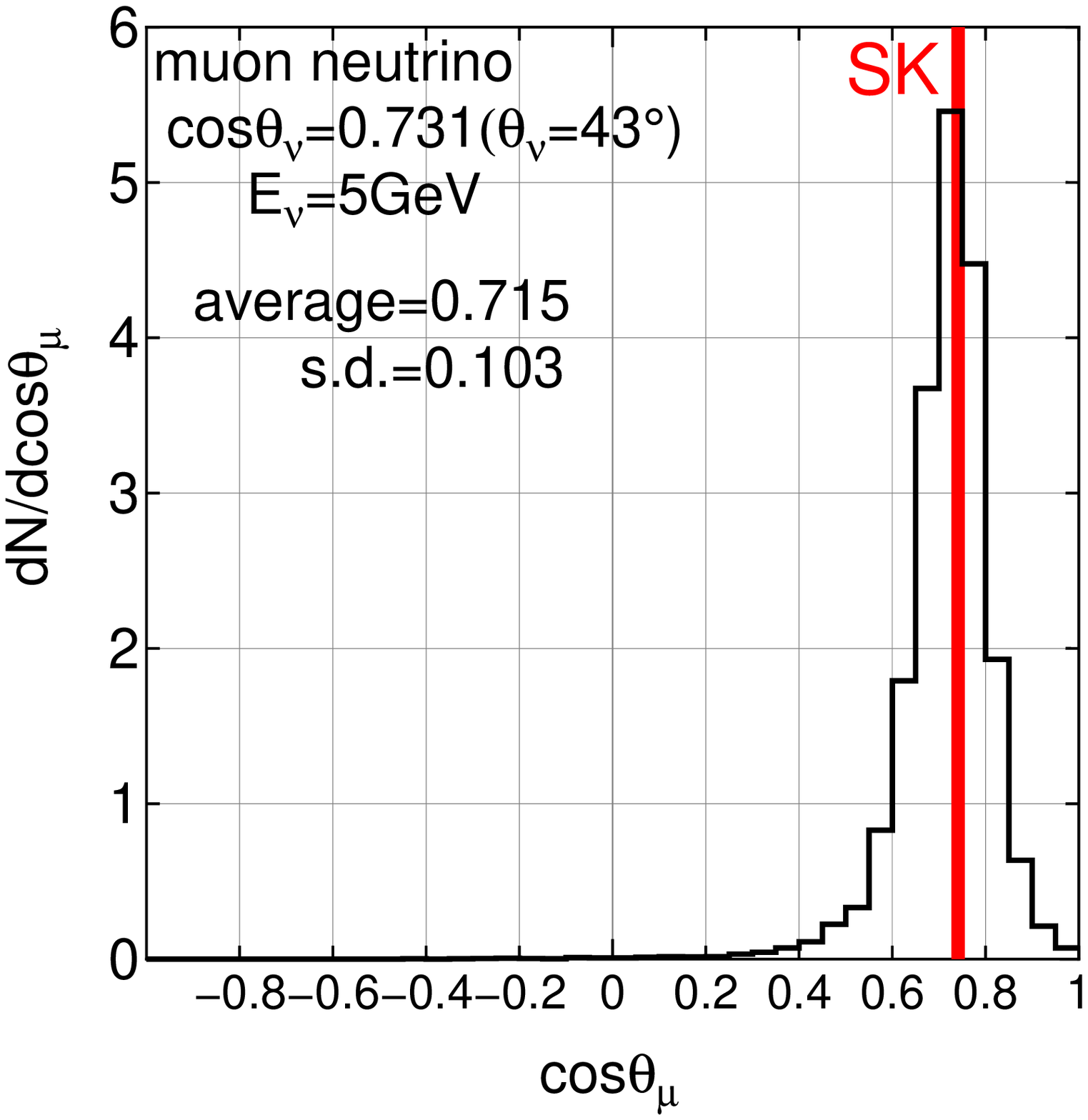}
  }
\caption{
\label{figH009} 
Zenith angle distribution of the muon for the diagonally incident muon 
neutrino with 0.5~GeV, 1~GeV and 5~GeV, respectively. The sampling 
number is 10000 for each case.
SK stands for the corresponding ones under the SK assumption.
}
\end{center}
\end{figure*} 

%
In Figures~\ref{figH007} to \ref{figH009}, 
we express Figures~\ref{figH004} to \ref{figH006} in a different way. 
We sum up muon events with different emitted energies for given 
zenith angles. As the result of it, we obtain frequency 
distribution of the neutrino events as  a function of 
$cos\theta_{\mu}$ for 
different incident directions and different incident energies of 
neutrinos.

In Figures~\ref{figH007}(a) to \ref{figH007}(c), we give the zenith angle 
distributions of the emitted muons for the case of vertically incident 
neutrinos with different energies, say,
 $E_{\nu}=$ 0.5, 1 and 5~GeV.

Comparing the case for 0.5 GeV with that for 5 GeV, we understand the big 
contrast between them as for the zenith angle distribution. The scattering 
angle of the emitted muon for 5 GeV neutrino is relatively small (See, 
Table 1), so that the emitted muons keep roughly the same direction as 
their 
original neutrinos. In this case, the effect of their azimuthal angle on 
the zenith angle is also smaller. However, in the case for 0.5 GeV 
which is the dominant energy for single ring muon events
in the Super-Kamiokande, there is even 
a possibility for the emitted muon to be emitted in the backward direction 
due to the larger angle scattering, the effect of which is enhanced by 
their azimuthal angle.

The most frequent occurrence in the backward direction of the emitted 
muon appears in the horizontally incident neutrino as shown in Figs. 8(a) 
to 8(c). In this case, the zenith angle distribution of the emitted muon 
should be symmetrical with regard
to the horizontal direction. Comparing the case for 
5 GeV with those both for $\sim$0.5 GeV and for $\sim$1 GeV, even 1 GeV 
incident neutrinos lose almost the original sense of the incidence if we 
measure it by the zenith angle of the emitted muon. 
Figures~\ref{figH009}(a) to \ref{figH009}(c) for 
the diagonally incident neutrinos tell us that the situation for diagonal 
case lies between the case for the vertically incident neutrinos and that 
for horizontally incident ones.
  SK in the figures denotes {\it the SK assumption on the direction} of 
incident neutrinos. 
From the Figures~\ref{figH007}(a) to \ref{figH009}(c), it is clear that the scattering 
angles of emitted muons influence their zenith angles, which is enhanced 
by their azimuthal angles, particularly for
more inclined directions of the incident neutrinos.

%
\section{Super-Kamiokande Assumption on the Direction
in the Light of $L_{\nu}$ and $L_{\mu}$}

In the previous section, we show that {\it the SK assumption on the direction} 
does not hold as for scattering angles of the leptons even if 
statistically.  This assumption is logically equivalent to the statement 
that $L_{\nu}$ is approximately the same as $L_{\mu}$ in $L/E$ analysis 
, where $L_{\nu}$ denotes the distance on the incident neutrino from 
the interaction point of the neutrino events to the intersection of 
the Earth surface toward its arriving direction and 
$L_{\mu}$ denotes the corresponding distance on the emitted muon.
Consequently, if our indication on the invalidity of
{\it the SK assumption on the direction} is correct,
 the same conclusion should be expected in the relation 
between $L_{\nu}$ and $L_{\mu}$.    
 In the present section and subsequent sections, we examine directly 
the validity of the implicit SK assumption that $L_{\nu}$ is 
approximated by $L_{\mu}$, 
taking into consideration the neutrino energy spectrum at the
Super-Kamiokande site.

\begin{figure}
\begin{center}
\vspace{-1cm}
\hspace*{-2.5cm}
\rotatebox{90}{%
\resizebox{0.5\textwidth}{!}{%
  \includegraphics{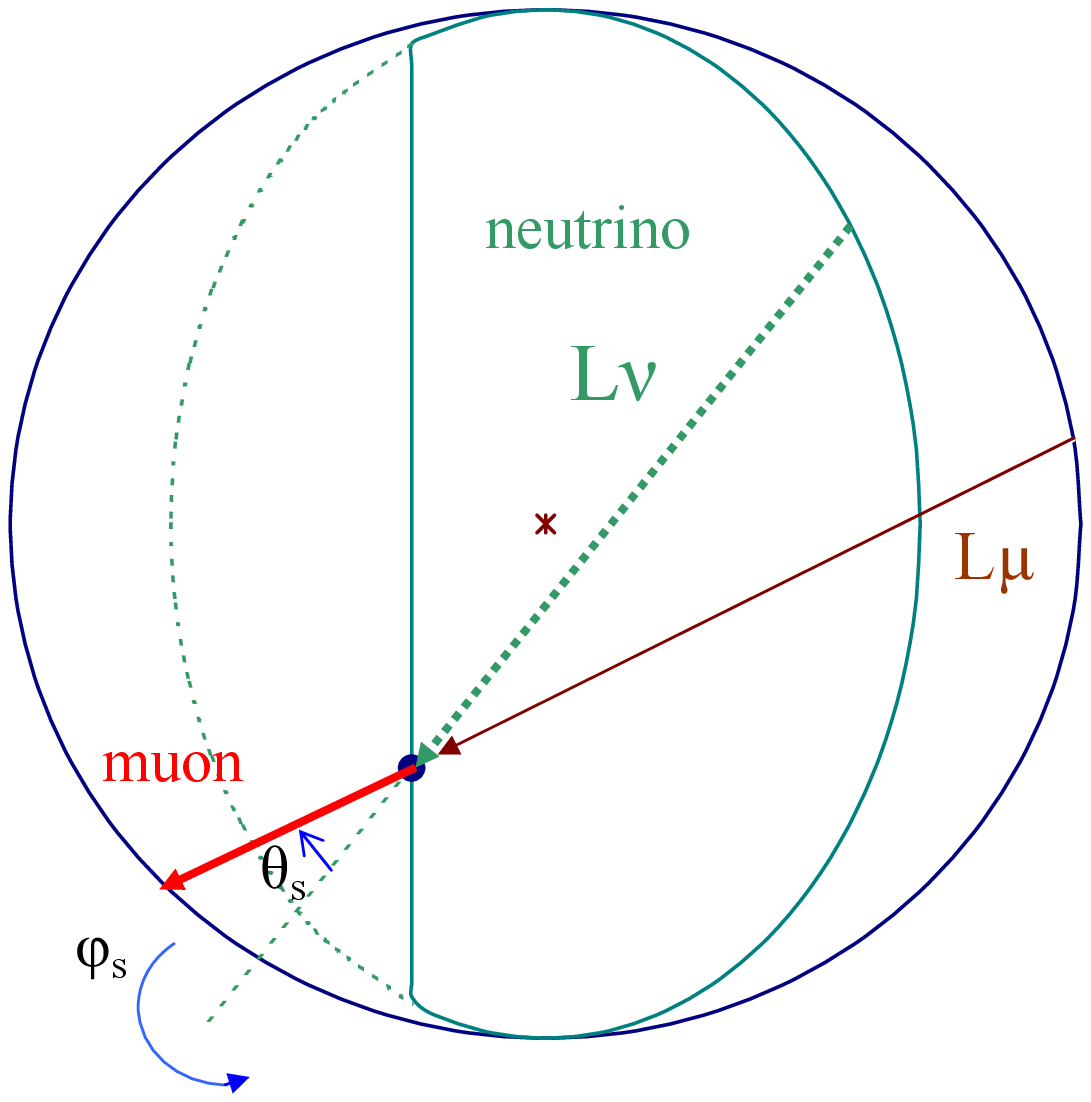}
}}
\vspace{-2cm}
\caption{Schematic view of relations among
$L_{\nu}$, $L_{\mu}$, $\theta_s$ and $\phi_s$
.}
\label{figH010}       
\resizebox{0.5\textwidth}{!}{%
  \includegraphics{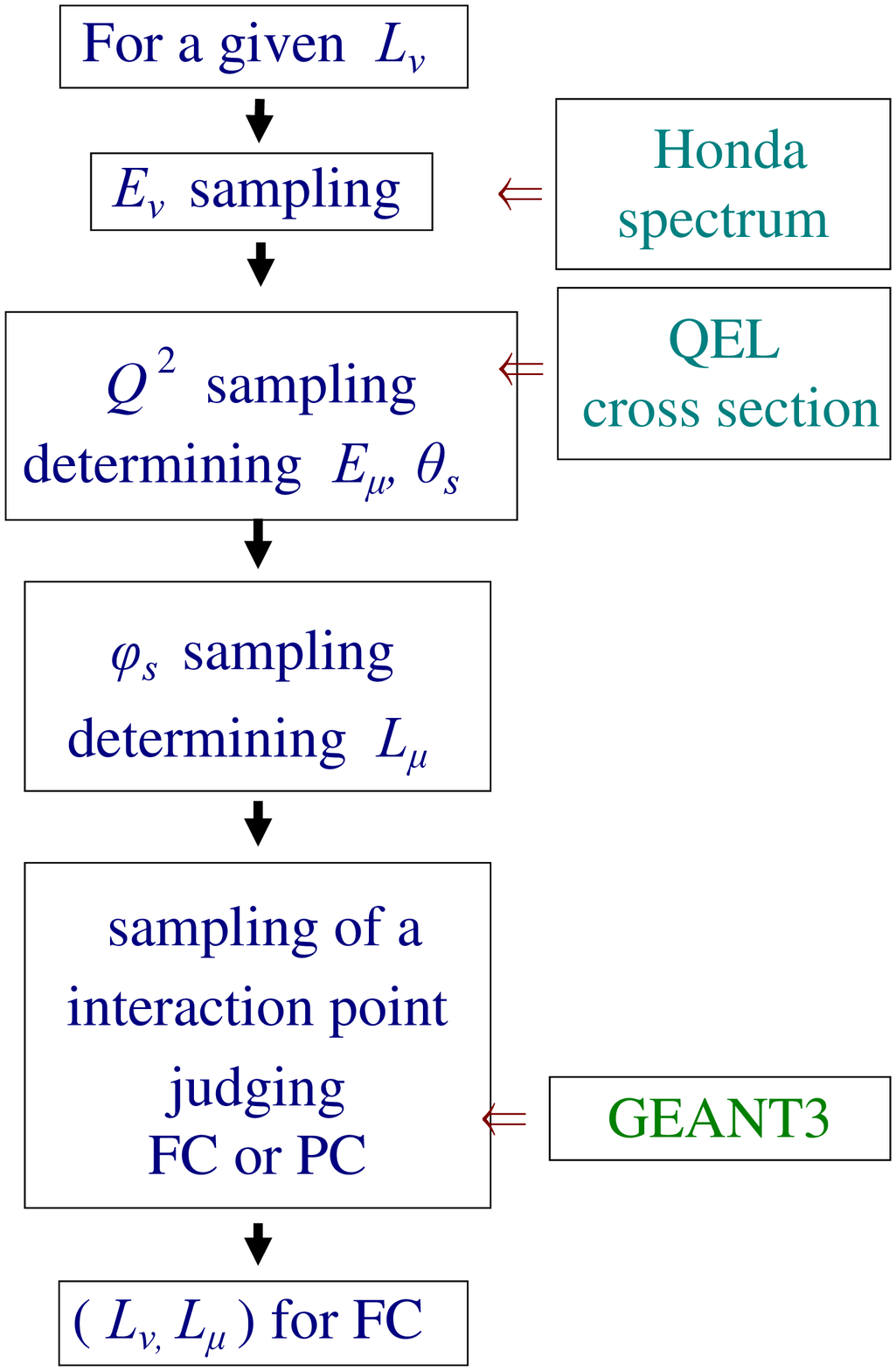}
}
\vspace{-1.5cm}
\caption{The procedure for our numerical experiment
for obtaining $L_{\mu}$ from a given $L_{\nu}$.
}
\label{figH011}
\end{center}
\end{figure}

\begin{figure}
\begin{center}
\vspace{-1cm}
\hspace*{-1.5cm}
\rotatebox{90}{%
\resizebox{0.5\textwidth}{!}{%
  \includegraphics{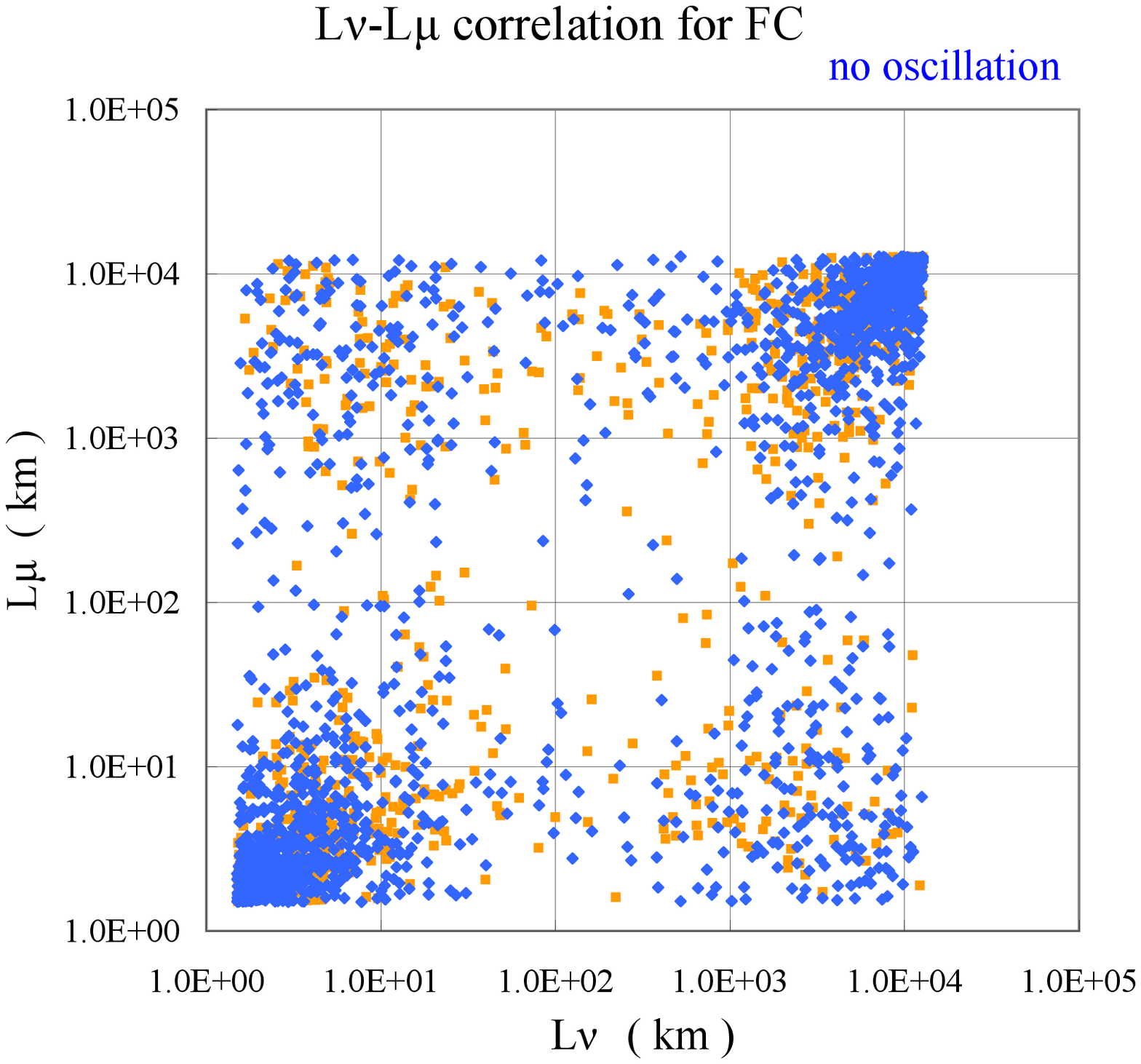}
}}
\vspace{-1.2cm}
\caption{Correlation diagram for $L_{\nu}$ and $L_{\mu}$ without 
oscillation for 1489.2 live days.
The blue points and orange points denote neutrino events and 
ani-neutrino events, respectively.  
}
\label{figH012}       
\vspace{-1cm}
\hspace*{-1.5cm}
\rotatebox{90}{%
\resizebox{0.5\textwidth}{!}{%
  \includegraphics{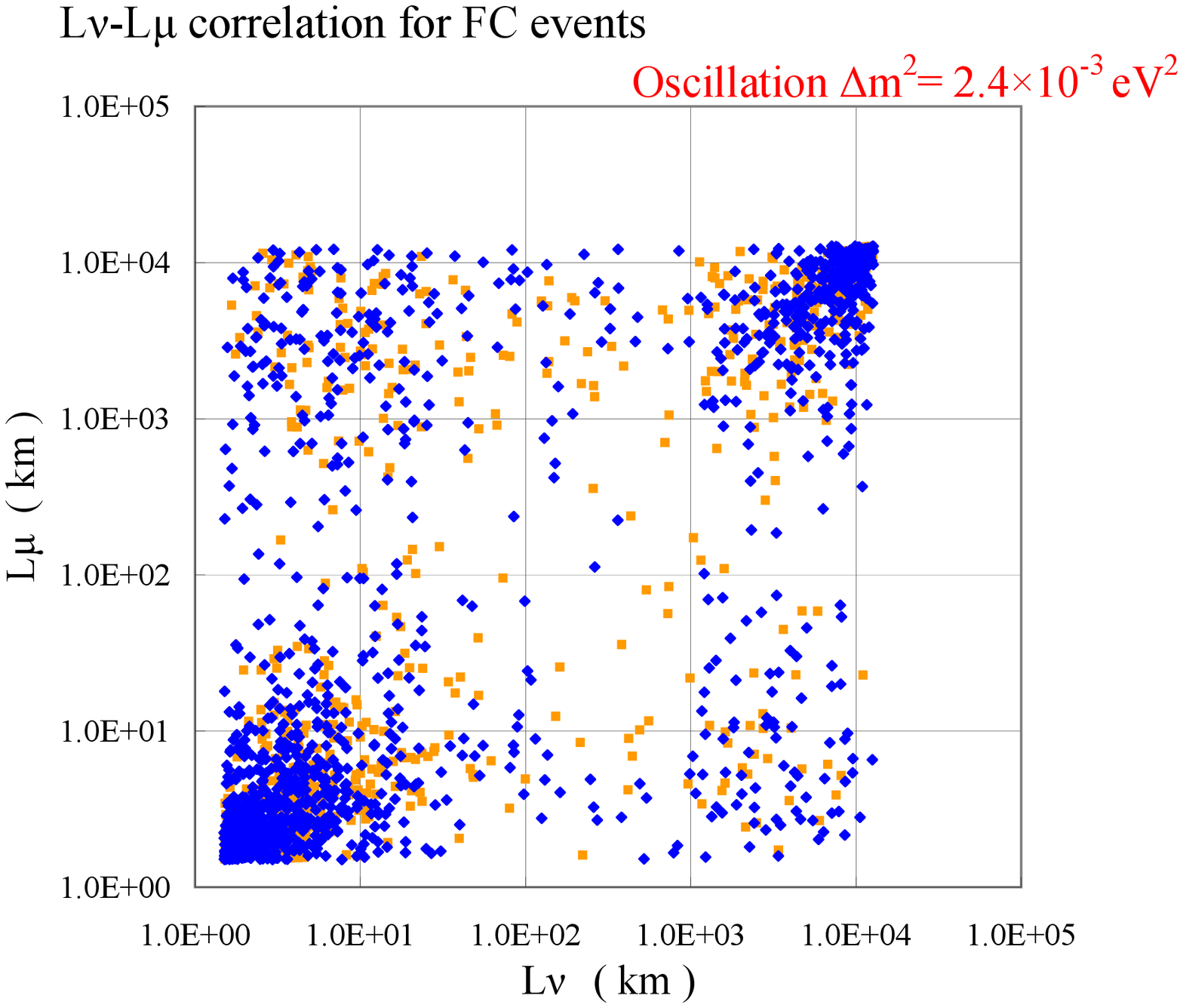}
}}
\vspace{-1.2cm}
\caption{Correlation diagram for $L_{\nu}$ and $L_{\mu}$ with 
 oscillation for 1489.2 live days.
The blue points and orange points denote neutrino events and 
ani-neutrino events, respectively.  
}
\label{figH013}
\end{center}
\end{figure}

In section~3.1 we give the procedure how to obtain $L_{\mu}$ from a 
neutrino event with given $L_{\nu}$ in the stochastic manner.
 In section~3.2 we give the correlations between $L_{\nu}$ and $L_{\mu}$, 
taking account of the effect of the backscattering as well as 
the effect of the azimuthal angle in the QEL in stochastic manner.
 As the result of it, we show that 
$L_{\nu}\approx L_{\mu}$,
 namely {\it the SK assumption on the direction}, 
does not hold even if statistically in both the absence and the 
presence of neutrino oscillation 
(Figure~\ref{figH012} and Figure~\ref{figH013}).
 Also, we treat the correlation between $E_{\nu}$ and $E_{\mu}$, 
in the stochastic manner. 
We show that the approximation of $E_{\nu}$ with $E_{\mu}$ 
by Super-Kamiokande Collaboration does not make so serious 
error compared with the approximation of $L_{\nu}$ by $L_{\mu}$,
 although their treatment is theoretically unsuitable 
(Figure~\ref{figH005}).

In section~4,
 we show that
$L_{\nu}/E_{\nu}$ distribution can reproduce the minimum extrema for 
neutrino oscillation which SK's
neutrino oscillation parameters demand and ,furthermore,
 it may give the differnt mimimum extrema in the neutrino 
oscillation under the different neutrino oscillation parameters from
SK's.
We show $L_{\nu}/E_{\nu}$ distribution can reproduce the minimum 
extrema for neutrino oscillation (the maximum oscillation) which 
Super-Kamiokande Collaboration demand, by using their neutrino 
oscillation parameters 
($\Delta m^2 = 2.4\times 10^{-3}\rm{ eV^2}$ and $sin^2 2\theta=1.0$).
 Furthermore, it may give the different minimum extrema in the neutrino
 oscillation under the different neutrino oscillation parameters from the 
Super-Kamiokande Collaboration.  
This fact denotes that our numerical computer experiment is done in 
a correct manner. 

\subsection{Derivation of $L_{\mu}$ from a given $L_{\nu}$ in the 
single ring muon event among Fully Contained Events} 

In our numerical computer experiment, we obtain single ring muon 
events among {\it the Fully Contained Events} resulting 
from QEL in the virtual SK detector,
 the details of which are described in Appendix A.
For the neutrino event with a definite neutrino energy thus generated,
we simulate its interaction point inside the detector and 
the emitted energy of the muon concerned 
which gives its scattering angle uniquely. 
The determination of the neutrino energy, the emitted energy of the muon 
and its scattering angle are described in Appendix A.
  The muon thus generated is pursued in the stochastic manner by using 
GEANT 3 and finally we judge whether the muon concerned stops inside the 
detector ({\it the Fully Contained Event}) or passes through 
the detector ({\it the Partially Contained Event}). 
 For {\it Fully Contained Events} thus obtained, we know the directions 
of the incident neutrinos, the generation points and termination points of 
the events generated inside the detector, the emitted muon energies,
their scattering angles and 
their azimuthal angles in QEL which give 
their zenith angles, $L_{\nu}$ and $L_{\mu}$
\footnote{
The azimuthal angle is but that in QEL, not that with
regard to the Earth here.}
finally.

 In Figure~\ref{figH010}, we show the relation between 
$L_{\nu}$ and $L_{\mu}$ schematically.
Figure~\ref{figH011} shows the procedure for obtaining $L_{\mu}$ from 
$L_{\nu}$ which is equivalent to the corresponding procedure for obtaining $cos\theta_{\mu}$ from $cos\theta_{\nu}$.

The relation between direction cosine of the incident neutrino, 
$(\ell_{\nu(\bar{\nu})}, m_{\nu(\bar{\nu})}, n_{\nu(\bar{\nu})} )$, and 
that of the corresponding emitted lepton, $(\ell_{\rm r}, m_{\rm r}, 
n_{\rm r})$, 
for a given scattering angle, $\theta_{\rm s}$,  
and its azimuthal angle, $\phi$, resulting from QEL is given in
Appendix A.

$L_{\nu}$ and $L_{\mu}$ are functions of the direction cosine of the 
incident neutrino, $cos\theta_{\nu}$, and that of emitted muon,
 $cos\theta_{\mu}$, respectively and they are given as,
$$
L_{\nu}= R_g \times ( r_{SK} cos\theta_{\nu} +
\sqrt{ r_{SK}^2 cos^2\theta_{\nu} + 1 - r_{SK}^2} )  \,\,\,\,(5-1)
$$
$$
L_{\mu}= R_g \times ( r_{SK} cos\theta_{\mu} +
\sqrt{ r_{SK}^2 cos^2\theta_{\mu} + 1 - r_{SK}^2} )  \,\,\,\,(5-2)
$$
\noindent where $R_g$ is the radius of the Earth and 
$r_{SK}=1-D_{SK}/R_g$, with the depth, $D_{SK}$, 
of the Super-Kamiokande
Experiment detector from the surface of the Earth.
It should be noticed that the $L_{\nu}$ and $L_{\mu}$ are regulated by
both the 
energy spectrum of the incident neutrino and the production spectrum of 
the muon (QEL in the present case). Consequently,
 their mutual relation is influenced by either 
the absence of the oscillation or the presence of the oscillation which 
depend on the combination of the oscillation parameters.

\begin{figure}
\begin{center}
\vspace{-1cm}
\hspace*{-1cm}
\rotatebox{90}{%
\resizebox{0.45\textwidth}{!}{%
  \includegraphics{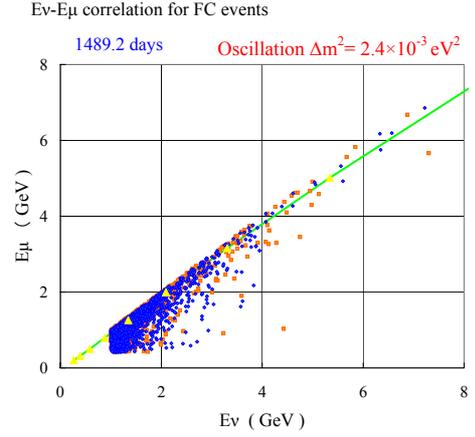}
}}
\vspace{-1.7cm}
\caption{The correlation diagram 
between $E_{\nu}$ and $E_{\mu}$ for 
oscillation for 1489.2 live days.  The continous line denotes 
the polynomial expression by Super-Kamiokande Collabolation.
}
\label{figH014}       
\end{center}
\end{figure}

\subsection{The correlation between $L_{\nu}$ and $L_{\mu}$} 


 In Figure~\ref{figH012}, we give the correlation diagram between
$L_{\nu}$ and $L_{\mu}$ for single ring muon events
among {\it Fully Contained Events}
 for the 1489.2 live days in the absence of neutrino oscillation
which corresponds to the actual Super-Kamiokande Experiment\cite{Ashie2}.  
In Figure~\ref{figH012}, blue points denote neutrino events while 
orange points denote anti-\\
neutrino events.
Throughout all correlation diagrams in the present paper,
blue points and orange ones have the same meaning shown in 
Figure~\ref{figH012}.
The aggregates of the (anti-) neutrino events which 
correspond to a definite 
combination of $L_{\nu}$ and $L_{\mu}$ are essentially classified into 
four groups in the following:

 Group A is defined as the aggregate for neutrino events in which 
both $L_{\nu}$ and $L_{\mu}$ are rather small. 
It denotes that the downward neutrinos produce the downward muons
with smaller scattering angles.
 In this case, the energies of the produced muons are near the 
energies of the incident neutrinos due to smaller scattering angles.

  Group B is defined as the aggregate for neutrino events in which 
both $L_{\nu}$ and $L_{\mu}$ are rather large.
It denotes that the upward neutrinos  produce upward muons
with smaller scattering angles.
In this case, the energy relation between the incident neutrinos and 
the produced muons is essentially the same as in Group A, because the 
flux of the upward neutrino events is symmetrical to that of the 
downward neutrino events in the absence of neutrino oscillation.

 Group C is defined as the aggregate for neutrino events in 
which $L_{\nu}$ are rather small and $L_{\mu}$ are rather large.
 It denotes that the downward neutrinos produce the 
upward muons by the possible effect reusulting from
 both backscattering and azimuthal angle in QEL.
 In this case, the energies of the produced muons are smaller 
than those of the energies of the incident neutrinos 
due to larger scattering angles.

 Group D is defined as the aggregate for the neutrino events in 
which $L_{\nu}$ are rather large and $L_{\mu}$ are rather small.
 It denotes that the upward neutrinos produce the 
downward muons. The energy relation between the incident neutrinos and 
the produced muons is essentially the same as in Group C in the absence 
of neutrino oscillation.

It is clear from Figure~\ref{figH012} that there exist the 
symmetries  between Group A and Group B, and also between Group C and 
Group D,  which reflect  
the symmetry between the upward neutrino flux and the downward neutrino 
one for null oscillation.  

 In Figure~\ref{figH013}, we give the correlation between 
$L_{\nu}$ and $L_{\mu}$
under their neutrino oscillation parameters, say,
$\Delta m^2 = 2.4\times 10^{-3}\rm{ eV^2}$ and $sin^2 2\theta=1.0$
\cite{Ashie2}. 
In the presence of neutrino oscillation, the property of the 
symmetry which holds in the absence of neutrino oscillation 
(see $\langle$Group A and Group B$\rangle$ and/or 
$\langle$Group C and Group D$\rangle$ in Figure~\ref{figH012})
is lost due to the different incident neutrino fluxes in the upward 
direction and downward one. 
If we compare Group A with Group B, the event 
number in Group B (upward $\nu$ $\rightarrow$ upward $\mu$)
is smaller than that in group A
(downward $\nu$ $\rightarrow$ downward $\mu$),
 which comes from smaller flux of the upward neutrinos. 
The similar relation between Group C 
(downward $\nu$ $\rightarrow$  upward $\mu$ ) and Group D 
(upward $\nu$ $\rightarrow$  downward $\mu$) is held in 
Figure~\ref{figH013}.

 Summarizing the characteristics among Groups A to D in the 
Figures~\ref{figH012} and \ref{figH013}, we could conclude that 
$\langle$Group A and Group B$\rangle$ and 
$\langle$Group C and Group D$\rangle$ are in symmetrical situations 
in the absence of neutrino oscillation,
while such a symmetrical situation is lost in the presence 
of neutrino oscillation.
  Also, it is clear from  Figures~\ref{figH012} 
and \ref{figH013} that $L_{\nu}\approx L_{\mu}$, namely  
{\it the SK assumption on the direction}, does not hold 
both in the absence of neutrino oscillation and 
in the presence of neutrino oscillation
even if statistically.

Here, it should be noticed that the approximation of
$L_{\nu}\approx L_{\mu}$ does not hold completely 
in the region C and region D.
The event numbers in Group C and Group D could not be neglected 
among the total event number concerned. 
 In these regions, neutrino events consist of those with 
backscattering and/or neutrino events 
in which the neutrino directions
are horizontally downward (upward),
 but their produced muons turn upward (downward) 
resulting from the effect of azimuthal angles in QEL.

\subsection{The correlation between $E_{\nu}$ and $E_{\mu}$}

 Super-Kamiokande Collaboration estimate $E_{\nu}$ from $E_{\mu}$,
 the visible energy of the muon,
 from their Monte Carlo simulation,
by the following equation\cite{Ishitsuka}
(see page 135 of the paper concerned) : 
$$
E_{\nu,SK}= E_{\mu}\times(a + b\times x + c\times x^2 + 
d\times x^3 ),  \,\,\,(6)
$$

\noindent where $x = log_{10}(E_{\mu})$.

 The idea that $E_{\nu}$ could be approximated as the polynomial means 
that there is unique relation between $E_{\nu}$ and $E_{\mu}$.
 However, in the light of stochastic characters inherent in both the 
incident neutrino energy spectrum and the production spectrum of 
the muon, such a treatment is not suitable theoretically,
 which may kill a real correlation effect between the incident 
neutrino energy and the emitted muon energy.
 In Figure~\ref{figH014}, we give the correlation between 
$E_{\nu}$ and $E_{\mu}$ 
together with that obtained from the polynomial expression by 
Super-Kamiokande Collaboration under 
their neutrino oscillation parameters 
and their incident neutrino energy spectrum\cite{Honda}. 
It is clear from the figure that the part of the lower energy 
incident neutrino deviates largely from the approximated formula,
 which reflects explicitly the stochastic character of QEL.
We can give the similar relation for null oscillation, the shape of 
which may be different from that with oscillation due to the 
difference in the incident neutrino energy spectrum.

Thus, we could choose four combinations, namely
$L_{\nu}/E_{\nu}$, $L_{\mu}/E_{\mu}$, $L_{\mu}/E_{\nu}$
and $L_{\nu}/E_{\mu}$
for the examination of maximum oscillations due to neutrino oscillation
in $L/E$ analysis. 
However, only the combination of  $L_{\mu}/E_{\mu}$ out of these four
combinations can be physically measurable.

\section{Summary}
Since one cannot measure $L_{\nu}$ and $E_{\nu}$, so one is forced to utilize
$L_{\mu}$ and $E_{\mu}$  
in the $L/E$ analysis in place of them.
 Then,  
Super-Kamiokande Collaboration 
assume that the direction of the incident neutrino is the same as 
that of the emitted lepton {\it the SK assumption on the direction} and
$E_{\nu}$ can be estimated from the some polynomial formula of the 
variable $E_{\mu}$ in $L/E$ analysis.
However, it is clear from Figures 12 and 13 that
 {\it the SK assumption on the direction}
 does not hold even approximately and the transformation of
$E_{\mu}$ into $E_{\nu}$ is not uniquely.

In the Part 2 of the subsequent paper, we apply the results from 
Figures 12, 13 and 14 to $L/E$ analysis and conclude that one cannot
 obtain the maximum oscillation in $L/E$ analysis which shows
 strongly the oscillation pattern from the neutrino oscillation.

\newpage
\noindent {\bf APPENDIX}\\
\appendix
\section{ 
 Monte Carlo Procedure for the Decision of Emitted Energies of the Leptons and Their Direction Cosines }
\setcounter {equation} {0}
\def\theequation{\Alph{section}\textperiodcentered\arabic{equation}}

Here, we give the Monte Carlo Simulation procedure for obtaining the energy and its  direction cosines,
$(l_{r},m_{r},n_{r})$, of the emitted lepton in QEL for a given energy and its direction cosines, $(l,m,n)$, of the incident neutrino. 

\begin{figure}
\begin{center}
\resizebox{0.45\textwidth}{!}{%
  \includegraphics{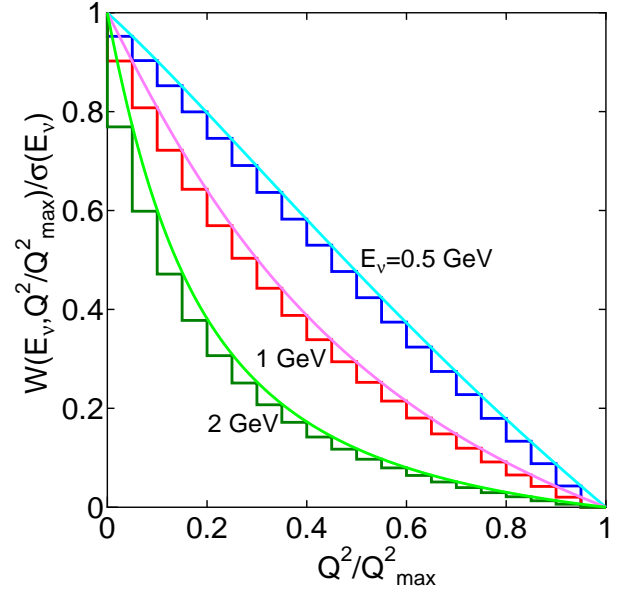}
  }
\end{center}
\caption{ The reappearance of the probability function for QEL cross section.
Histograms are  sampling results, while the curves  concerned are theoretical
ones for given incident energies.
}
\label{figP001}
\end{figure}

The relation among $Q^2$, $E_{\nu(\bar{\nu})}$, the energy of the incident neutrino, $E_{\ell(\bar{\ell})}$, the energy of the emitted lepton (muon or electron or their anti-particles) and $\theta_{\rm s}$, the scattering angle of the emitted lepton, is given as
      \begin{equation}
         Q^2 = 2E_{\nu(\bar{\nu})}E_{\ell(\bar{\ell})}(1-{\rm cos}\theta_{\rm s}).
\label{eqn:a1}  
      \end{equation}
\noindent Also, the energy of the emitted lepton is given by
      \begin{equation}
         E_{\ell(\bar{\ell})} = E_{\nu(\bar{\nu})} - \frac{Q^2}{2M}.
\label{eqn:a2}  
      \end{equation}
\noindent {\bf Procedure 1}\\
\noindent
We decide  $Q^2$ from the probability function for the differential cross section with a given $E_{\nu(\bar{\nu})}$ (Eq. (\ref{eqn:2}) in the text) by using the uniform random number, ${\xi}$,  between (0,1) in the following\\
  \begin{equation}
    \xi = \int_{Q_{\rm min}^2}^{Q^2}P_{\ell(\bar{\ell})}(E_{\nu(\bar{\nu})},Q^2)
                             {\rm d}Q^2,
\label{eqn:a3}  
  \end{equation}
\noindent where
  \begin{eqnarray}
\lefteqn{     P_{\ell(\bar{\ell})}(E_{\nu(\bar{\nu})},Q^2) =} \nonumber \\
&&  \frac{ {\rm d}\sigma_{\ell(\bar{\ell})}(E_{\nu(\bar{\nu})},Q^2) }{{\rm d}Q^2} 
                     \Bigg /\!\!\!\!
      \int_{Q_{\rm min}^2}^{Q_{\rm max}^2} 
      \frac{ {\rm d}\sigma_{\ell(\bar{\ell})}(E_{\nu(\bar{\nu})},Q^2) }{{\rm d}Q^2} 
             {\rm d}Q^2 . \nonumber \\
&&
\label{eqn:a4}  
   \end{eqnarray}
\\
\noindent From Eq. (A$\cdot$1), we obtain $Q^2$ in histograms together with the corresponding theoretical curve in Figure~\ref{figP001}. The agreement between the sampling data and the theoretical curve is excellent, which shows the validity of the utlized  procedure in Eq. (A$\cdot$3) is right. \\

\begin{figure}
\begin{center}
\resizebox{0.45\textwidth}{!}{%
  \includegraphics{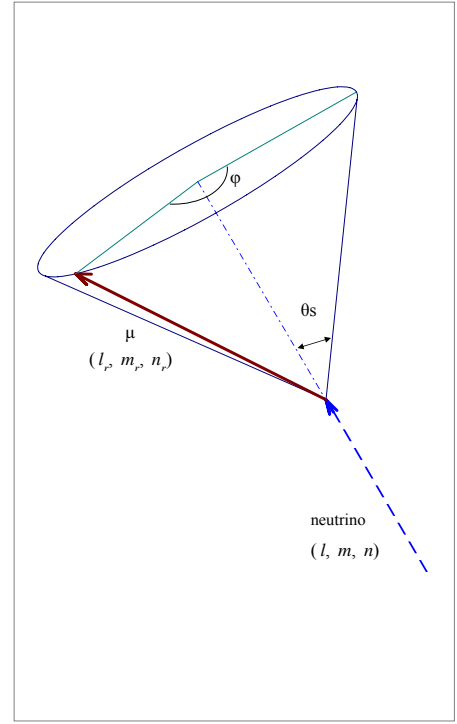}
  }
\end{center}
\caption{ The relation between the direction cosine of the incident neutrino and that of the emitted charged lepton.}
\label{figP002}
\end{figure}

\noindent {\bf Procedure 2}\\
\noindent
We obtain $E_{\ell(\bar{\ell})}$ from Eq. (A$\cdot$2) for  the given $E_{\nu(\bar{\nu})}$ and $Q^2$ thus decided in the Procedure 1.\\

\noindent {\bf Procedure 3}\\
\noindent
We obtain $\cos{\theta_{\rm s}}$, cosine of the the scattering angle of the emitted lepton, for $E_{\ell(\bar{\ell})}$ thus decided in the Procedure 2 from Eq. (A$\cdot$1) .\\

\noindent {\bf Procedure 4}\\
\noindent
We decide $\phi$, the azimuthal angle of the scattering lepton, which is obtained from\\
  \begin{equation}
       \phi = 2\pi\xi.                 
\label{eqn:a5}  
  \end{equation}
\noindent Here, $\xi$ is a uniform random number between (0, 1). \\
As explained schematically in the text(see Figure~{\bf \ref{figH003}} 
in the text),  we must take account of the effect due to the azimuthal 
angle $\phi$ in the QEL to obtain the zenith angle distribution both
for {\it Fully Contained Events} and {\it Partially Contained Events} 
correctly.\\  

\noindent {\bf Procedure 5}\\
\noindent
The relation between direction cosines of the incident neutrinos, $(\ell_{\nu(\bar{\nu})}, m_{\nu(\bar{\nu})}, n_{\nu(\bar{\nu})} )$, and those of the corresponding emitted lepton, $(\ell_{\rm r}, m_{\rm r}, n_{\rm r})$, for a certain $\theta_{\rm s}$ and $\phi$ is given as
\begin{equation}
\left(
         \begin{array}{c}
             \ell_{\rm r} \\
             m_{\rm r} \\
             n_{\rm r}
         \end{array}
       \right)
           =
       \left(
         \begin{array}{ccc}
           \displaystyle \frac{\ell n}{\sqrt{\ell^2+m^2}} & 
            -\displaystyle 
            \frac{m}{\sqrt{\ell^2+m^2}}        & \ell_{\nu(\bar{\nu})} \\
            \displaystyle \frac{mn}{\sqrt{\ell^2+m^2}} & \displaystyle 
            \frac{\ell}{\sqrt{\ell^2+m^2}}     & m_{\nu(\bar{\nu})}    \\
                        -\sqrt{\ell^2+m^2} & 0 & n_{\nu(\bar{\nu})}
         \end{array}
       \right)
       \left(
          \begin{array}{c}
            {\rm sin}\theta_{\rm s}{\rm cos}\phi \\
            {\rm sin}\theta_{\rm s}{\rm sin}\phi \\
            {\rm cos}\theta_{\rm s}
          \end{array}
       \right),
\label{eqn:a6}
\end{equation}

\noindent where $n_{\nu(\bar{\nu})}={\rm cos}\theta_{\nu(\bar{\nu})}$,
 and $n_{\rm r}={\rm cos}\theta_{\ell}$. 
Here, $\theta_{\ell}$ is the zenith angle of the emitted lepton. \\

The Monte Carlo procedure for the determination of $\theta_{\ell}$ of the emitted lepton for the parent (anti-)neutrino with given $\theta_{\nu(\bar{\nu})}$ and $E_{\nu(\bar{\nu})}$ involves the following steps:\\

We obtain $(\ell_r, m_r, n_r)$ by using Eq. (\ref{eqn:a6}). The $n_r$ is the cosine of the zenith angle of the emitted lepton which should be contrasted to $n_{\nu}$, that of the incident neutrino.
\\
Repeating the procedures 1 to 5 just mentioned above, we obtain the zenith angle distribution of the emitted leptons for a given zenth angle of the incident neutrino with a definite energy. \\

In the SK analysis,  instead of Eq. (\ref{eqn:a6}), they assume \\ $n_r = n_{\nu(\bar{\nu})} $ 
uniquely for ${E_{\mu(\bar{\mu})}} \geq$ 400 MeV.\\

%

\end{document}